\documentclass[aps,pre,10pt,superscriptaddress,footinbib,longbibliography,twocolumn,amsmath,amssymb,final,letterpaper]{revtex4-1}
\usepackage{graphicx}
\usepackage{subfigure}
\usepackage{times}
\usepackage{bm}
\usepackage{xcolor}
\usepackage{marginnote} 
\usepackage{mathrsfs}

\usepackage[colorlinks=true,linkcolor=blue,urlcolor=blue,citecolor=blue,anchorcolor=blue]{hyperref}

\hypersetup{pdfdisplaydoctitle=true,pdftitle={Spectral dimension reduction of complex dynamical networks}}
\hypersetup{pdfauthor={E. Laurence, N. Doyon, L.J. Dubé, and P. Desrosiers}}
\pacs{}

\newcommand{\obs}{{}}
\newcommand{\toperi}{s_{pc}} 
\newcommand{\tocore}{s_{cp}} 
\newcommand{\spectralradius}{r}
\newcommand{\binaryedge}{B}

\begin{document}

\title{Spectral dimension reduction of complex dynamical networks}

\author{Edward Laurence}
\affiliation{D\'epartement de physique, de g\'enie physique, et d'optique, Universit\'e Laval, Qu\'ebec, G1V 0A6, Canada}
\affiliation{Centre interdisciplinaire en mod\'elisation math\'ematique de l'Universit\'e Laval, Qu\'ebec, G1V 0A6, Canada}
\author{Nicolas Doyon}
\affiliation{Centre interdisciplinaire en mod\'elisation math\'ematique de l'Universit\'e Laval, Qu\'ebec, G1V 0A6, Canada}
\affiliation{D\'epartement de math\'ematiques et de statistique, Universite Laval, Qu\'ebec, G1V 0A6, Canada}
\affiliation{Centre de recherche CERVO, Qu\'ebec, G1J 2G3, Canada}
\author{Louis J. Dub\'e}
\affiliation{D\'epartement de physique, de g\'enie physique, et d'optique, Universit\'e Laval, Qu\'ebec, G1V 0A6, Canada}
\affiliation{Centre interdisciplinaire en mod\'elisation math\'ematique de l'Universit\'e Laval, Qu\'ebec, G1V 0A6, Canada}
\author{Patrick Desrosiers}
\affiliation{D\'epartement de physique, de g\'enie physique, et d'optique, Universit\'e Laval, Qu\'ebec, G1V 0A6, Canada}
\affiliation{Centre interdisciplinaire en mod\'elisation math\'ematique de l'Universit\'e Laval, Qu\'ebec, G1V 0A6, Canada}
\affiliation{Centre de recherche CERVO, Qu\'ebec, G1J 2G3, Canada}

\begin{abstract} 
Dynamical networks are powerful tools for modeling a broad range of complex systems, including financial markets, brains, and ecosystems. They encode how the basic elements (nodes) of these systems interact altogether (via links) and evolve (nodes' dynamics). Despite substantial progress, little is known about why some subtle changes in the network structure, at the so-called critical points, can provoke drastic shifts in its dynamics. We tackle this challenging problem by introducing a method that reduces any network to a simplified low-dimensional version. It can then be used to describe the collective dynamics of the original system. This dimension reduction method relies on spectral graph theory and, more specifically, on the dominant eigenvalues and eigenvectors of the network adjacency matrix. Contrary to previous approaches, our method is able to predict the multiple activation of modular networks as well as the critical points of random networks with arbitrary degree distributions. Our results are of both fundamental and practical interest, as they offer a novel framework to relate the structure of networks to their dynamics and to study the resilience of complex systems.
\end{abstract}
%==============================================================================
\maketitle

%==============================================================================
\section{Introduction}
\label{sec:introduction}
%==============================================================================
Critical breakdowns generally arise unexpectedly in complex dynamical systems \cite{scheffer2009critical}. Noteworthy examples are financial crises \cite{schweitzer2009economic,gualdi2015tipping}, epileptic seizures \cite{richardson2012large}, and species extinctions \cite{sole2001complexity}. These breakdowns are typically identified by using global-scale indicators that collapse at the critical point, such as stock market indices, neural synchronization, and species biomass. While much effort has been devoted to forecast breakdowns \cite{boettiger2013tipping}, no simple and universal method has yet been found. This is mostly due to the inherent complexity of the problem: real systems are composed of multiple units that participate to the global state in highly complicated patterns of interactions.   

Network Science addresses this problem and offers a unifying framework where a complex system with $N$ fundamental units is described as a network of $N$ components (nodes). The state of each unit is encoded into an activity variable and the evolution of the states in the whole system is governed by $N$ coupled dynamical equations that depend on both the activity variables and a set of weighted interactions (links). Therefore, the dynamical properties of the system strongly depend on the underlying network structure. Although recent development has clarified how small targeted perturbations in the network structure can provoke drastic changes in the structure itself \cite{morone2015influence,lu2016vital}, much less is known about the dynamical effects of these perturbations. One promising approach is to use dimension reduction to transform the original $N$--dimensional representation into a simplified version with $n\ll N$ effective dimensions. 

Recently, Gao \textit{et al.} have presented a dimension reduction formalism that collapses any $N$--dimensional network into a $1$--dimensional effective version and used it to predict the global activity of the original network \cite{gao2016universal}. The authors have proposed to measure the global activity as the degree-weighted average activity in which the nodes with high degree, i.e., high number of links, contribute more to the average than those with low degree. The rationale behind this choice is that the highly connected nodes have a higher impact on the dynamics. Moreover, they have shown that the degree-weighted connectivity is sufficient to explain the global level of activity. Their formalism can be applied to a wide variety of complex systems, thus suggesting that the degree-weighted averages are in fact \textit{universal} predictors. For instance, they accurately predict the minimum level of interaction between species to prevent biomass extinction, the so-called \textit{critical point}, of real ecosystems. Yet, this spectacular outcome is not totally satisfactory since no fundamental reason is provided that would explain why degree is the key property to any network structure, particularly those with degree correlations.

In an attempt to determine the critical points of 59 bipartite mutualistic ecosystems \cite{jiang2018predicting}, Jiang \textit{et al.} have proposed a $2$--dimensional reduction that divide each original ecosystem into two populations for which they obtain the average interaction strength. From numerical explorations, they conclude that the degree may not always be the key predictive property of a network. Their results also suggest that $2$--dimensional reductions can lead to better predictions than the $1$--dimensional formalism of Gao \textit{et al.}.

It remains unclear whether the dimension reduction procedure of Gao \textit{et al.} can lead to accurate predictions for arbitrary network structures and why some dynamical networks should require 2--dimensional reductions. Besides, strong theoretical foundations are clearly lacking to answer those questions. The goal of this paper is to address these issues. We rely on a simple and strong hypothesis: one can predict the evolution of a small number of variables describing the global activities of a network. These variables are constructed as a priori unknown weighted-averages of the individual node activities. By enforcing this hypothesis, we provide theoretical justifications for the required number of effective dimensions and quantify the contribution of each component to the universal global activities. Beyond the mere improvement in precision over existing approaches, our method allows the detection of dynamical breakdowns that would be missed altogether with previous reductions.

This study provides a reliable tool for researchers who want to study critical breakdowns of complex systems. Using our approach, once a system is framed as a network, one can first determine the number of effective dimensions required to adequately reduce the system at hand, and then find the variables of interest that describe the global state. Moreover, our method also identifies the units that, if perturbed, can induce large reactions in the system. Our findings thus lead to a deeper understanding of how critical breakdowns occur and how to prevent them.

The paper is structured as follows. We first present the general framework for complex dynamical networks  (Sec.~\ref{subsec:modeldefinition}). We then describe a general method to obtain a 1--dimensional reduction for mutualistic networks (Sec.~\ref{sec:effective_system}). We show that the reduction scheme of Gao \textit{et al.} emerges as an approximation of our general approach when specifically considering random networks. In Sec.~\ref{subsec:multidimensional} we develop the \textit{cycle reduction}, a multidimensional approach useful to reduce heterogeneous and bipartite networks. Next, in Sec.~\ref{sec:linear combinaison}, we complete the method by including subdominant contributions of the structure. We finally assess the goodness of these reductions, as a function of the structure, and the nature of the dynamics (Sec.~\ref{subsec:scalefree}).

%==============================================================================
\section{Model definition}
\label{subsec:modeldefinition}
%==============================================================================

The diverse nature of complex systems requires to establish a common ground. In Sec.~\ref{subsec:generalformalism}, we regroup dynamical complex networks under a general model that encodes the structure and the dynamics. Then, in Sec.~\ref{subsec:one:examples}, we provide examples of contrasting models of dynamics satisfying the formalism used afterward in the paper to illustrate the dimension reduction methods.

%==============================================================================
\subsection{General formalism}
\label{subsec:generalformalism}
%==============================================================================

We consider a complex network of $N$ units, called \textit{nodes}, for which the interactions are encoded in the weighted and directed adjacency matrix $\bm{W}$. The element $w_{ij}\in \mathbb{R}$ of $\bm{W}$ is interpreted as the strength of the directed interaction from node $j$ to node $i$.

Each node has an activity $x_i\in\mathbb{R}$ whose evolution is governed by the general equation
\begin{equation}
  \dot{x}_i = F(x_i) + \sum_{j=1}^N w_{ij}G(x_i,x_j) \label{Eq:complete_dynamics},
\end{equation}
where $F(x_i)$ and $G(x_i,x_j)$ are real-valued functions. For technical reasons that will become clear in the next section, both $F(x_i)$ and $G(x_i,x_j)$ are required to have continuous derivatives of second order. The product $w_{ij}G(x_i,x_j)$ specifies the type of interactions. 

If $w_{ij}\partial G(x_i,x_j)/\partial x_j\leq 0$, the interaction is competitive and the increase of activity of node $j$ tends to decrease the activity of node $i$. If $w_{ij}\partial G(x_i,x_j)/\partial x_j\geq0$, the interaction is mutualistic and therefore node $j$ activity benefits node $i$. For mixed dynamics of mutualistic and competitive interactions, it is common to fix $\partial G(x_i,x_j)/\partial x_j\geq0$ and use negative weights $w_{ij}<0$ for competitive interactions.
Unless specified, we will only consider mutualistic dynamics with $w_{ij}\partial G(x_i,x_j)/\partial x_j\geq 0$. 
Furthermore, we have concentrated our studies of the possible dynamical models (Sec.~\ref{subsec:one:examples}) in parameter ranges for which $\partial G(x_i,x_j)/\partial x_j\geq 0$ such that $w_{ij}$ are all non-negative as well.

To describe the evolution of the whole system at the macro- and the mesoscopic scales, it is convenient to focus on \textit{observables}. We define an observable as a \textit{smooth} function mapping the activities $x_1,\ldots, x_N$ to a real number. Among all observables, the linear observables, functions of the form $L(\bm x)=\sum_ia_ix_i\in\mathbb{R}$, are of particular interest for different reasons. 

The first reason to use a linear observable is the numerical evidence given by Ref.~\cite{gao2016universal} that suggests that a linear combination of the activity could be a good indicator of the global state of the network. We have followed this direction, which has led us to a more general linear dimension reduction formalism. 
The second reason is that a linear combination is far more intuitive than a general non-linear observable of the form $\Omega(\bm{x})  = \sum_{n,j}  a_{nj} x_j^n$.
 It is not excluded that such non-linear observables could provide adequate dimension reductions, but their interpretability would most surely be limited.
Finally, there is a more formal and practical reason to select linear observables.  The set of all observables, which are smooth functions from $\mathbb{R}^N$ to $\mathbb{R}$, forms a real vector space $\mathscr{V}$ of infinite dimension.  Dimension reduction can thus be seen as a search for $n<N$ observables, among infinitely many possible choices,  whose evolution can be well approximated by $n$ coupled differential equations. At first sight, there is no obvious way to determine the best observables to choose.  However, $\mathscr{V}$ contains a subspace $\mathscr{L}$ formed by all linear observables.  The dimension of $\mathscr{L}$ is precisely $N$ and one can show that any basis of $\mathscr{L}$ has a dynamics that is exactly described by $N$ differential equations very similar to the original ones.  This simplifies the search immensely:  rather than looking in the whole space  $\mathscr{V}$, the spectral properties of the adjacency matrix to find $n$ elements of interest in $\mathscr{L}$ offer themselves as a natural choice.

%==============================================================================
\subsection{Examples of possible dynamics}
\label{subsec:one:examples}
%==============================================================================

\begin{subequations}
\label{Eqs:examples}
A number of dynamical systems satisfy the form of Eq.~\eqref{Eq:complete_dynamics}. For instance, in computational neuroscience, the \emph{Cowan-Wilson model} \cite{wilson1972excitatory} describes the firing-rate activity of a population of neurons as
\begin{equation}
	\dot{x}_i = -x_i + \sum_{j=1}^N w_{ij}\dfrac{1}{1+\exp[-\tau(x_j-\mu)]}\label{eq:cowan-wilson},
\end{equation}
where $\tau$ and $\mu$ are parameters controlling the steepness of the activation function and the firing-rate threshold, respectively \footnote{This is actually a modified version of the original Cowan-Wilson model, which describes output activities of the specific form of \unexpanded{Eq.~\eqref{Eq:complete_dynamics}}. In Eq.\unexpanded{~\eqref{eq:cowan-wilson}}, \unexpanded{$x_i$} must be interpreted as the input activity to neuron  \unexpanded{$i$}.}.

In biology, the \emph{generalized Lotka-Volterra dynamics} describes the evolution of the population of species in an ecosystem as
\begin{equation}
 \dot{x}_i = \omega x_i + x_i\sum_{j=1}^N w_{ij}x_j,\label{eq:lotka-volterra}
\end{equation}
where $\omega$ is the intrinsic growth rate \cite{grilli2017feasibility}, and $x_i$ is the population of individuals of species $i$. To prevent unbounded growth and account for species migration and the {\em Allee effect}, a more complex model of ecological networks has been proposed \cite{holland2002population,gao2016universal}:
\begin{align}
	\dot{x}_i= B_i+&x_i\left(1-\dfrac{x_i}{K_i}\right)\left(\dfrac{x_i}{C_i}-1\right)\nonumber\\
	&+\sum_{j=1}^N w_{ij}\dfrac{x_ix_j}{D_i+E_ix_i+H_ix_j},\label{Eq:ecosystem:complex}
\end{align}
where all parameters are real-valued, $B_i$ accounts for the migration rate, $K_i>0$ for the ecosystem capacity, and $C_i> 0$ for the minimum abundance for species growth. The parameters $D_i,E_i,H_i$ control the strengths of the interactions between the species.

The \emph{Michaelis-Menten equation} is yet another example \cite{alon2006introduction}.  It applies to the gene regulatory networks and governs the concentration of substrates as
\begin{equation}
	\dot{x}_i=-cx^a_i+\sum_{j=1}^N w_{ij}\dfrac{x^b_j}{x^b_j+1},\label{eq:michaelis-menten}
\end{equation}
where $a,b,c\in\mathbb{R}$ are parameters. 

In social networks, the spreading of a virus or rumors can be described using the \emph{Susceptible-Infected-Susceptible model} (SIS) \cite{pastor2001epidemic}. In this context, the activity $x_i\in[0,1]$ is interpreted as the probability of being infected and evolves according to
\begin{equation}
	\dot{x}_i=-x_i+\gamma(1-x_i)\sum_{j=1}^Nw_{ij}x_j\label{eq:sis}, 
\end{equation}
with $\gamma\geq 0$ as the normalized infection rate. 
\end{subequations}

%==============================================================================
\section{1--dimensional reduction}
\label{sec:effective_system}
%==============================================================================

The systems described by Eqs.~\eqref{Eqs:examples} are $N$--dimensional, their dynamics governed by $N$ coupled differential equations. As the number of nodes $N$ grows, the computational cost of solving $N$ coupled equations increases which raises a number of issues \cite{eliasmith2012large,laurence2018exact}. Moreover, the state of the original system given by the $N$--dimensional vector $\bm{x}$ becomes less intelligible and less insightful, and does not provide much into the general properties of the solutions.

Hence, we must rely on measures, or observables, to reduce $N$--dimensional systems to more practical and accessible objects. For instance, the unweighted average activity could be a   measure on how dissimilar the system state is compared to a specifically chosen state. We also want to make predictions on those measures to anticipate dynamical breakdowns and locate the global state of the system on a standardized bifurcation diagram. However, when solely based on the unweighted average activity, the predictions are often non-representative of the original system \cite{jiang2018predicting}. 

Alternatively, a \textit{weighted} activity seems more reliable as we inject additional information on the importance of nodes and has already been proven to be a promising avenue of breakdown predictions \cite{gao2016universal,jiang2018predicting}. In the next subsections, we introduce a general procedure to select a weighted  activity and to predict its evolution.

\begin{figure}[t!]
  \centering
  \includegraphics[width=\linewidth]{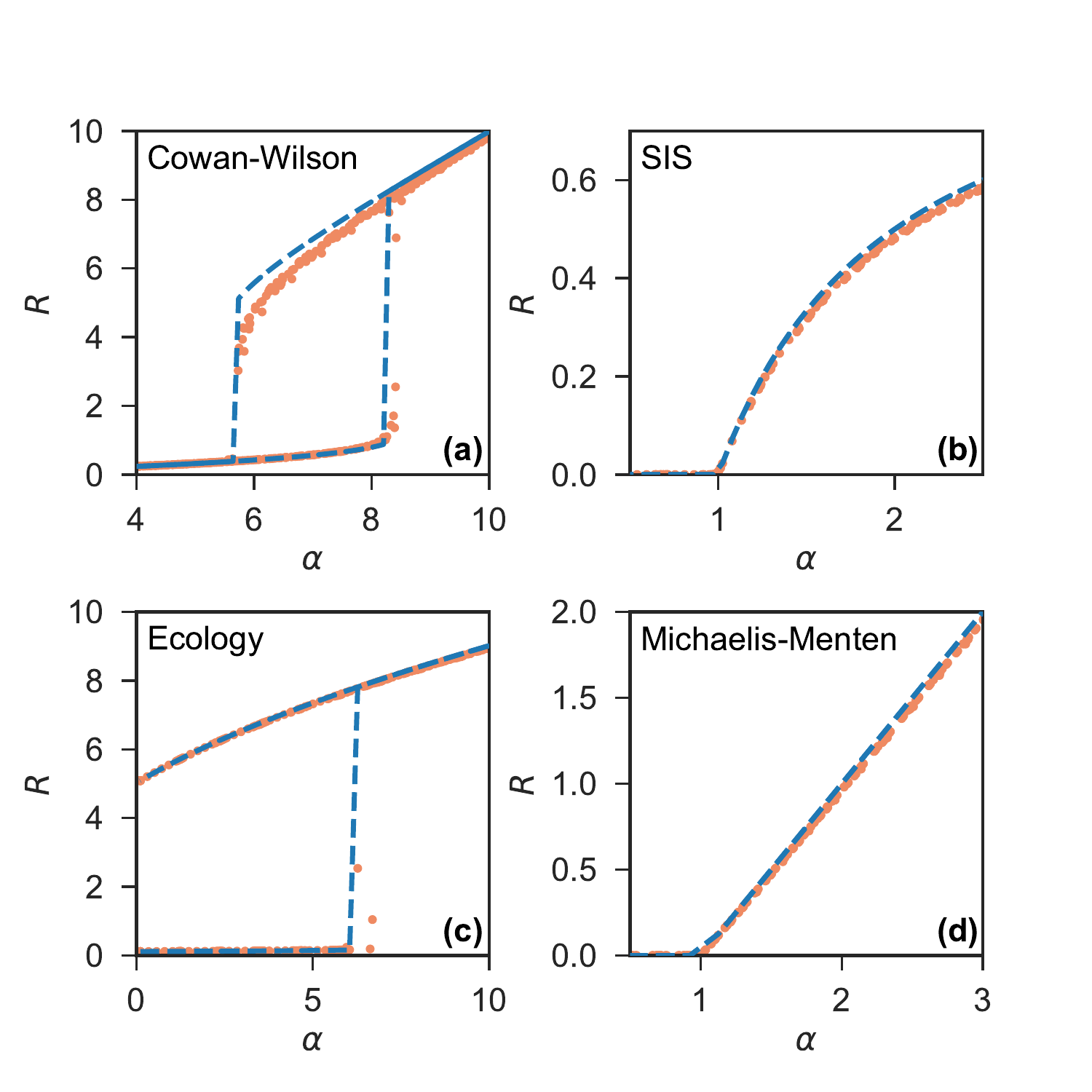}
  \caption{(Color online) Observable $R^*=\bm{a}^T\bm{x}^*$ at equilibrium as a function of the dominant eigenvalue $\alpha$ of $\bm{W}^T$, for different dynamics on Erd\H{o}s-R\'{e}nyi networks of $N=100$ nodes and
connection probability $p=0.1$. (a) Cowan-Wilson dynamics Eq.~\eqref{eq:cowan-wilson} with $\tau=1, \mu=3$, (b) SIS dynamics Eq.~\eqref{eq:sis} with $\gamma=1$, (c) Mutualistic ecological dynamics Eq.~\eqref{Eq:ecosystem:complex} with $B_i=0.1,C_i=1,K_i=5,D_i=6,E_i=0.9,H_i=0.1$ \cite{barabasi1999emergence}, (d) Michaelis-Menten dynamics Eq.~\eqref{eq:michaelis-menten} with $a=1,b=1,c=1$. Dashed lines are theoretical predictions obtained from Eq.~\eqref{eq:uni:equilibirum} while dots are equilibrium states resulting from the evolution of the whole $N$--dimensional system. For each dynamics and network ensemble, 100 networks are generated. For each network, we scale the edge weights by a constant random factor as $w_{ij}\mapsto cw_{ij}$, so that the dominant eigenvalue of the adjacency matrix is located in the region of interest. Then, the dynamics are integrated to equilibrium and an orange dot is placed at the corresponding point $(\alpha,R^*)$. Next, the network is perturbed by removing an edge and the dynamics is brought back to equilibrium, and a new dot at $(\alpha',R'^*)$ is placed. The perturbation step is repeated 50 times for each network. }
  \label{fig:multid}
\end{figure}

%==============================================================================
\subsection{Derivation of the reduction formalism}
\label{subsec:1--dimensional}
%==============================================================================

Let us consider a real linear observable $R$ of the activity:
\begin{equation}
  R = \sum_{i=1}^N a_i x_i=\bm{a}^T\bm{x},\label{eq:R}
\end{equation} 
where $a_i\in\mathbb{R}$, the $i$-th component of the column vector $\bm{a}$, is a normalized weight so that
\begin{equation}
	\sum_{i=1}^N a_i=\bm{1}^T\bm{a}=1.
\end{equation}
In general, some components $a_i$ can be positive or negative, and $R$ represents a weighted activity. In many instances however, $a_i$ will be non-negative, i.e. the normalized vector $\bm{a}$ 
will be a probability vector, and our  observable $R$ could then be called justifiably a weighted {\em average} activity, where $a_i$ is the relative contribution, or \textit{centrality}, of node $i$ to the observable.

The linear observable $R$ is a function that takes the instantaneous activity of each node and returns a real number that describes the global state of the network. For instance,  for $a_i=1/N$, $R$ describes the unweighted average activity. Although the average activity is attractive because of its simplicity, it may not be easy to predict its value using only the structure of the network and the nature of the dynamics. Thus, we hypothesize that $\bm{a}$ should be specific to the structure. 

Let us explain how the weight vector $\bm{a}$ is constrained by the adjacency matrix $\bm{W}$. By taking the time derivative of Eq.~\eqref{eq:R} and using Eq.~\eqref{Eq:complete_dynamics} (refer to Appendix \ref{sec:app:derivation} for complete derivation), we obtain that the dynamics of $R$ -- truncated up to second-order terms $O[(x_k-R)^2]$~--  is given by the 1--dimensional equation 
\begin{equation}
  \dot{R} \approx F(R)+ \alpha G(\beta R, R)\label{eq:R_dynamics},
\end{equation}
where $\beta$ is a structural parameter given by
\begin{equation}
  \beta = \dfrac{1}{\alpha }\dfrac{\bm{a}^T\bm{K}\bm{a}}{\bm{a}^T\bm{a}}\label{Eq:uni:beta}
\end{equation}
and $\bm{K}$ is a $N\times N$ diagonal matrix of diagonal elements $K_{ii}=k_i^{\text{in}}=\sum_{j=1}^N w_{ij}$, the in-degree of node $i$. 
The parameter $\alpha$ can be measured directly on the network as the weighted in-degree,
\begin{equation}
  \alpha = \sum_{i=1}^N a_i k_i^{\text{in}}=\bm{a}^T\bm{k}^{\text{in}}.\label{Eq:uni:alpha}
\end{equation}
Interestingly, we show, in Appendix \ref{sec:app:derivation}, that the closed form of Eq.~\eqref{eq:R_dynamics} is satisfied only if $\bm{a}$ is a normalized eigenvector of the transposed adjacency matrix $\bm{W}^T$
with eigenvalue $\alpha$,  
\begin{equation}
 \bm{W}^T \bm{a}= \alpha \bm{a}.
\end{equation}

We now have obtained a single equation [Eq.~\eqref{eq:R_dynamics}] that governs the evolution of the weighted activity of a complex network, and constrained the weight vector $\bm{a}$ to be adapted to the structure under study. 

Clearly, Eq.~\eqref{eq:R_dynamics} shares similarities with Eq.~\eqref{Eq:complete_dynamics}. We can interpret the former as a reduced system of one dynamical node that interacts with itself. The nature of its dynamics is identical to the one from the original system, i.e. specified by $F(x_i)$ and $G(x_i,x_j)$, and the coupling is parametrized by $\alpha, \beta$. The activity $R$ of the single node describes the weighted activity of the original network, and is obtained by solving Eq.~\eqref{eq:R_dynamics} which solutions are solely controlled by the nature of the dynamics and the structural parameters $\alpha, \beta$.

%==============================================================================
\subsection{Choice of a universal weight vector}
\label{subsec:weightvector}
%==============================================================================

We have seen that the weight vector $\bm{a}$ must be a normalized eigenvector of $\bm{W}^T$ so that the observable $R$ satisfies Eq.~\eqref{eq:R_dynamics}. In principle, any eigenvectors of $\bm{W}^T$, except those that satisfy $\bm{1}^T\bm{a}=0$, could be used for the dynamical reduction. However, the larger the modulus of $\alpha$ is, the stronger is the influence of the structure on the weighted activity. An eigenvector whose eigenvalue has a low modulus leads to a linear observable $R$ that does not properly take into account the network structure, which in turn leads to correction terms $O[(x_k-R)^2]$ greater than those produced by eigenvectors with a higher modulus. 
The choice of $\bm{a}$ as the eigenvector with the largest eigenvalue modulus seems to impose itself: $\bm{a}$ is the dominant eigenvector.

For an arbitrary weighted adjacency matrix, the dominant eigenvalue and the components of the dominant eigenvector can be complex. In this case, the observable $R$ as well as the structural parameters $\alpha$ and $\beta$ are complex. The 1--dimensional dynamical system of Eq.~\eqref{eq:R_dynamics} becomes complex too and can be interpreted as a 2--dimensional real dynamical system.

There is however a large class of networks for which the dominant eigenvalue and the components of the dominant eigenvector are all real. For instance, strongly connected (in practice, sufficiently connected) undirected and directed networks with non-negative weights $w_{ij}$ and 
 fall into this class. 
In fact, the Perron-Frobenius theorem guarantees that if the network is strongly connected, i.e. a path exists between each pair of nodes, and that all edge weights satisfy $w_{ij}\geq 0$, then the dominant eigenvalue $\lambda_D$ of $\bm{W}^T$ is non-negative $\lambda_D\geq 0$, and the dominant eigenvector is elementwise positive~\cite{horn1990matrix}. Moreover, in practice, the dominant eigenvector can be efficiently computed using the power method. 

The procedure to apply this 1--dimensional dimension reduction is straightforward. First, we compute the dominant eigenvalue $\alpha$ and the corresponding eigenvector $\bm{v}_D$ of $\bm{W}^T$. Second, we define the normalized eigenvector $\bm{a}=\bm{v}_D/(\bm{1}^T\bm{v}_D)$, and obtain $\beta$ according to Eq.~\eqref{Eq:uni:beta}. In most cases, we want to determine the weighted activity at equilibrium  $R^*$, determined by solving
\begin{equation}
	0=F(R^*)+\alpha G(\beta R^*, R^*).\label{eq:uni:equilibirum}
\end{equation}
This is a universal equation in the sense that $\alpha$ and $\beta$ are independent of the dynamics, controlled by $F(x_i)$ and $G(x_i,x_j)$; $\alpha$ and $\beta$ only depend upon the network structure, encoded in $\bm{W}$. 
The 1--dimensional reduction process has been applied to different dynamics for small random uncorrelated networks and led to surprisingly accurate predictions (Fig.~\ref{fig:multid}). For larger networks, we expect the formalism to maintain a similar level of accuracy. However, since the number of nodes does not intervene explicitly in the formalism, we are generally unable to analytically describe how the quality of the reduction varies with the network size. We must rely on a numerical investigation.

Our numerical experiments indicate that the network size {\em by itself} has no significant impact on the quality of the reduction. Rather, the accuracy strongly depends on the network structure and, in particular, on the degree variance. 
In a nutshell, our findings can be summarized as follows:\\
a. For a given $N$, the larger the average \textit{number of edges per node} $\langle \bm{k}_{\binaryedge}^{in}\rangle$\footnote{The average number of edges per node \unexpanded{$\langle \bm{k}_{\binaryedge}^{in}\rangle=N^{-1}\langle\bm{1}^T\bm{A}\bm{1}\rangle$} is computed from the binary adjacency matrix \unexpanded{$\bm{A}$} and must be distinguished from the average degree \unexpanded{$\langle \bm{k}^{in}\rangle=N^{-1}\langle\bm{1}^T\bm{W}\bm{1}\rangle$} computed from the weighted adjacency matrix \unexpanded{$\bm{W}$}. In the latter equations, \unexpanded{$\langle\bm{1}^T\bm{A}\bm{1}\rangle$} either denotes the expected value of \unexpanded{$\bm{1}^T\bm{A}\bm{1}=\sum_{i,j}A_{ij}$} over a random network ensemble (in the case of random networks such as ER networks) or simply \unexpanded{$\sum_{i,j} A_{ij}$} (in the case of a particular network), and similarly for \unexpanded{$\langle\bm{1}^T\bm{W}\bm{1}\rangle$}.}, the better the dimensional reduction will be; \\
b. For a fixed value of $\langle \bm{k}^{in}_{\binaryedge} \rangle$, the residual error of the reduction tends rapidly to a finite limit as $N$ is increased;\\
c. For large enough $\langle \bm{k}^{in}_{\binaryedge} \rangle$,  the reduction error is small and insensitive to the precise value of $N$ (Corollary of a. and b.).

\begin{figure}
		\centering
		\includegraphics[width=\linewidth]{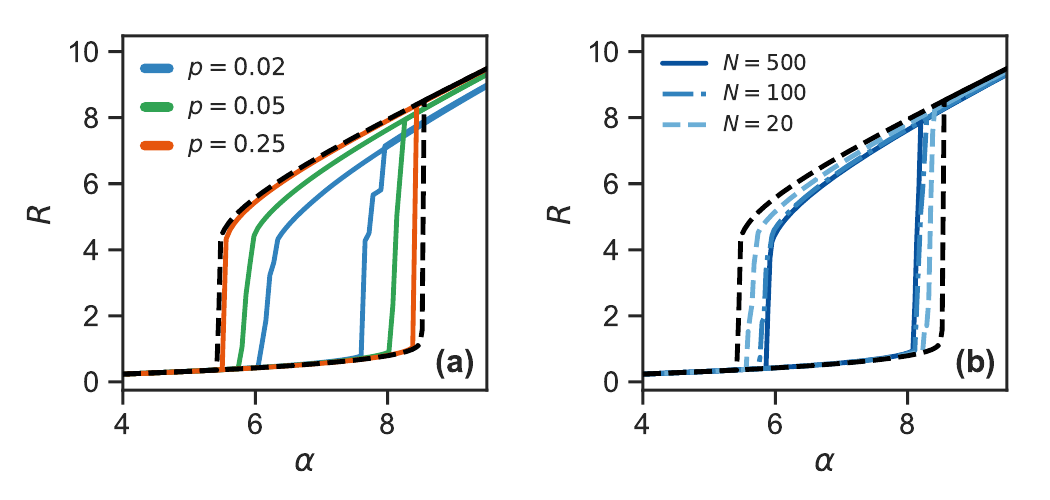}
		\caption{(Color online)
		(a) Comparison of the observable $R^*=\bm{a}^T\bm{x}^*$ at equilibrium as a function of the dominant eigenvalue $\alpha$ of $\bm{W}^T$ for different connection probabilities $p$ of undirected Erd\H{o}s-Rényi networks $G(N,p)$ and the Cowan-Wilson dynamics Eq.~\eqref{eq:cowan-wilson} with $\tau=1, \mu=3$. For each value of $p$, 10 networks of 200 nodes are generated and their dynamics are solved for $\alpha\in[4,10]$. Lines are computed as a binned average over same connection probabilty networks.
(b) Comparison of Erd\H{o}s-Rényi networks $G(N,p)$ of different sizes $N\in\lbrace 20, 100, 500 \rbrace$ but equal average number of edges per node $\langle \bm{k}^{in}_{\binaryedge}\rangle=10$. For each $N$, we adjust the connection probabilities $p \in \lbrace 0.50, 0.10, 0.02 \rbrace$ to match the expected value of $\langle \bm{k}^{in}_{\binaryedge}\rangle$ and solve the dynamics on 10 network realizations. Lines are computed as a binned average over same size networks. Dashed lines are theoretical predictions obtained from Eq.~\eqref{eq:uni:equilibirum}. To obtain a specific $\alpha= \bm{a}^T \bm{k}^{in}$, we multiply each edge weight by a constant scaling factor $w_{ij}\mapsto c w_{ij}$ so that the dominant eigenvalue $\alpha \mapsto c \alpha$ falls in the range $[4,10]$ using the dominant eigenvector $\bm{a}$ as weight vector. We then solve the dynamics at equilibrium and measure the observable $R^*=\bm{a}^T\bm{x}^*$. }
		\label{Fig2.pdf}
\end{figure}

In Fig.~\ref{Fig2.pdf}, we examine these conclusions by comparing different network ensembles. Although displayed for undirected Erd\H{o}s-Rényi (ER) networks,  $G(N,p)$, the results are representative of a larger set of calculations, and synthesize nicely our general conclusions on this issue.
In Fig.~\ref{Fig2.pdf}(a), dynamics on networks of different $\langle \bm{k}^{in}_{\binaryedge} \rangle= p(N-1) $
but equal number of nodes $N=200$, are differently reproduced by the 1--dimensional reduction. Denser networks (large $p$, large $\langle \bm{k}^{in}_{\binaryedge} \rangle$) are better represented by the reduction than sparser 
(small $p$, small $\langle \bm{k}^{in}_{\binaryedge} \rangle$) networks. 
In Fig.~\ref{Fig2.pdf}(b), we compare networks of different sizes but equal $\langle \bm{k}^{in}_{\binaryedge}\rangle$. For a fixed $\langle \bm{k}^{in}_{\binaryedge} \rangle$  (here $\langle \bm{k}^{in}_{\binaryedge} \rangle = 10$), 
the connection probability $p$ is adjusted to compensate for the growing number of nodes.  
As the number of nodes is increased, the residual error on the prediction rapidly tends to a finite limit, and no further deterioration of the quality of the dimensional reduction is observed. In other words, the goodness 
of the reduction is practically invariant of the network size, if large enough, and is mostly governed by the average number of edges per node. 

These observations extent to other types of networks and confirm that the quality of the reduction is more affected by the network connectivity than by the network size. A dynamical explanation goes as follows. In sparse networks (e.g. $N=200, p= 0.02$), one must use large scaling factors  $c  > 1$ ($\bm{W} \mapsto c\bm{W}$ which implies  $\bm{W}\bm{1}=\bm{k}^{in} \mapsto c \bm{k}^{in}$, $\alpha= \bm{a}^T \bm{k}^{in} \mapsto c \alpha$) to reach the desired range of the dominant eigenvalue $\alpha\in[4,10]$ (see caption of Fig.~\ref{Fig2.pdf}: the scaling does not alter the topology, only the strength of what is injected in the dynamical equations through the parameter $\alpha$). 
This accentuates the inequalities between the degrees of the nodes, $\text{Var}(\bm{k}^{in}) \mapsto c^2 \, \text{Var}(\bm{k}^{in})$, and eventually splits the populations into active and inactive nodes. Hence, the reduction is unable to describe the two populations with a single variable. In the opposite extreme of dense networks (e.g. $N=200, p=0.25$), the edge weights must be scaled down, $c < 1$,  to reach the same dominant eigenvalue. Therefore, the nodes follow a more global activation scheme that can be described with high accuracy by the 1--dimensional reduction. This explanation matches the observed finite limit for the quality of the reduction [Fig.~\ref{Fig2.pdf}(b)]. For a fixed value of $\langle \bm{k}^{in}_{\binaryedge} \rangle= p(N-1)$, when $N$ grows larger, $p$ tends to zero so that the variance $\text{Var}(\bm{k}_{\binaryedge}^{in})\approx(N-1)p(1-p)$ tends to a limit $\text{Var}(\bm{k}_{\binaryedge}^{in})\to \langle \bm{k}_{\binaryedge}^{in} \rangle$ and no further deterioration of the quality is observed. 

The quantitative impact of the degree variance is best explained by a simple example. Consider an ER network, ($N=200, p=0.1)$ with $\langle \bm{k}_{\binaryedge}^{in}\rangle\approx 20$ and $c=1$ so that the variance of the edge weights is 
$\text{Var}(\bm{k}^{in}) \approx 18$ and the dominant eigenvalue is $\alpha\approx 20$. For a denser network ($N=200, p=0.5$) with  $\langle \bm{k}_{\binaryedge}^{in}\rangle\approx 100$,

one must use a small scaling factor,  $\bm{W}\mapsto\bm{W}/5$, to have the same dominant eigenvalue $\alpha\approx 20$, and the scaled variance of the edge weights is now $\text{Var}(\bm{k}^{in}) \approx 2$. Therefore, the nodes in the denser networks are more uniformly activated than in the sparser networks and are better amenable to the 1--dimensional reduction.

%==============================================================================
\subsection{Choice of an approximate weight vector}
\label{subsec:randomnetworks}
%==============================================================================

Recently, Gao \textit{et al.} \cite{gao2016universal} have introduced a different 1--dimensional  reduction for dynamics of the form of Eq.~\eqref{Eq:complete_dynamics}. In this section, we show how their reduction is  a special case of our 1--dimensional reduction when applied to uncorrelated random networks.

Uncorrelated random networks are a family of networks for which the degree distribution can be arbitrary but the probability of connection between two nodes is independent of the presence or absence of any other edge~\cite{newman2001random}.
We generate our random networks using the configuration model \cite{newman2018networks}. We first sample the nodes in- and out- expected degrees $\bm{\kappa}^{\text{in}}, \bm{\kappa}^{\text{out}}$ from an arbitrary degree distribution. Then, we connect node $j$ to node $i$ with probability
\begin{equation}
  P_{ij}=\dfrac{\kappa^{\text{in}}_i\kappa^{\text{out}}_j}{m}\label{Eq:random_condition},
\end{equation}
where $m=\sum_i \kappa^{\text{out}}_i$ is the expected total number of edges. If the resulting network is strongly connected, the Perron-Frobenius theorem guarantees that the dominant eigenvector $\bm{v}_D$ of $\bm{W}^T$ will have only non-negative elements. We may then use this dominant eigenvector to construct the observable 
\begin{equation}
	R=\bm{a}^T\bm{x}=\dfrac{\bm{v}^T_D\bm{x}}{\bm{1}^T\bm{v}_D}.
\end{equation}
For networks that satisfy Eq.~\eqref{Eq:random_condition}, spectral graph theory \cite{chung2003spectra} informs us that the elements of  
the dominant eigenvector $\bm{v}_D$ of $\bm{W}^T$ (i.e. the weights of the reduced system)  can be approximated by the vector of out-degrees $\bm{k}^{\text{out}}$ as 
\begin{equation}
  a_i = [\bm{v}_D]_i \approx \dfrac{k^{\text{out}}_i}{\sum_{j=1}^N k^{\text{out}}_j}.
\end{equation}
if the rather mild condition 
\begin{equation}
	\frac{\langle (k^{\text{out}})^2\rangle}{\langle k^{\text{out}}\rangle}>\sqrt{\max[k_i^{\text{out}}]}\ln(N) , \label{Eq:condition_random}
\end{equation}
is satisfied. It then results from Eq.~\eqref{Eq:uni:alpha} that $\alpha$ measures the average neighbor in-degree, that is
\begin{subequations}
\begin{equation}
  \alpha \approx \dfrac{\sum_{i=1}^N k^{\text{out}}_i k^{\text{in}}_i}{\sum_{i=1}^N k^{\text{out}}_i},
\end{equation}
 and Eq.~\eqref{Eq:uni:beta} reduces to $\beta=1$. Therefore, $R$ is simply the average neighbor activity:
\begin{equation}
  R \approx \dfrac{\sum_{i=1}^N k^{\text{out}}_i x_i}{\sum_{i=1}^N k^{\text{out}}_i}.
\end{equation}
\end{subequations}
It turns out that this special case is exactly the formalism proposed by Gao \textit{et al.}\cite{gao2016universal} with $R=x_{\text{eff}}$ and $\alpha=\beta_{\text{eff}}$, in their notation.  It also means that the formalism of Gao \textit{et al.} is mostly appropriate for random networks [Eq.~\eqref{Eq:random_condition}] respecting Eq.~\eqref{Eq:condition_random}. Moreover, a recent work \cite{castellano2017relating} has introduced a corrected eigenvalue approximation for random networks with power-law degree distribution $p(k)\sim k^{-\gamma}$ with $\gamma>5/2$. This may further limit the accuracy of Gao \textit{et al.} approach with respect to our 1--dimensional reduction scheme.

\begin{figure}[t!]
  \centering
  \includegraphics[width=\linewidth]{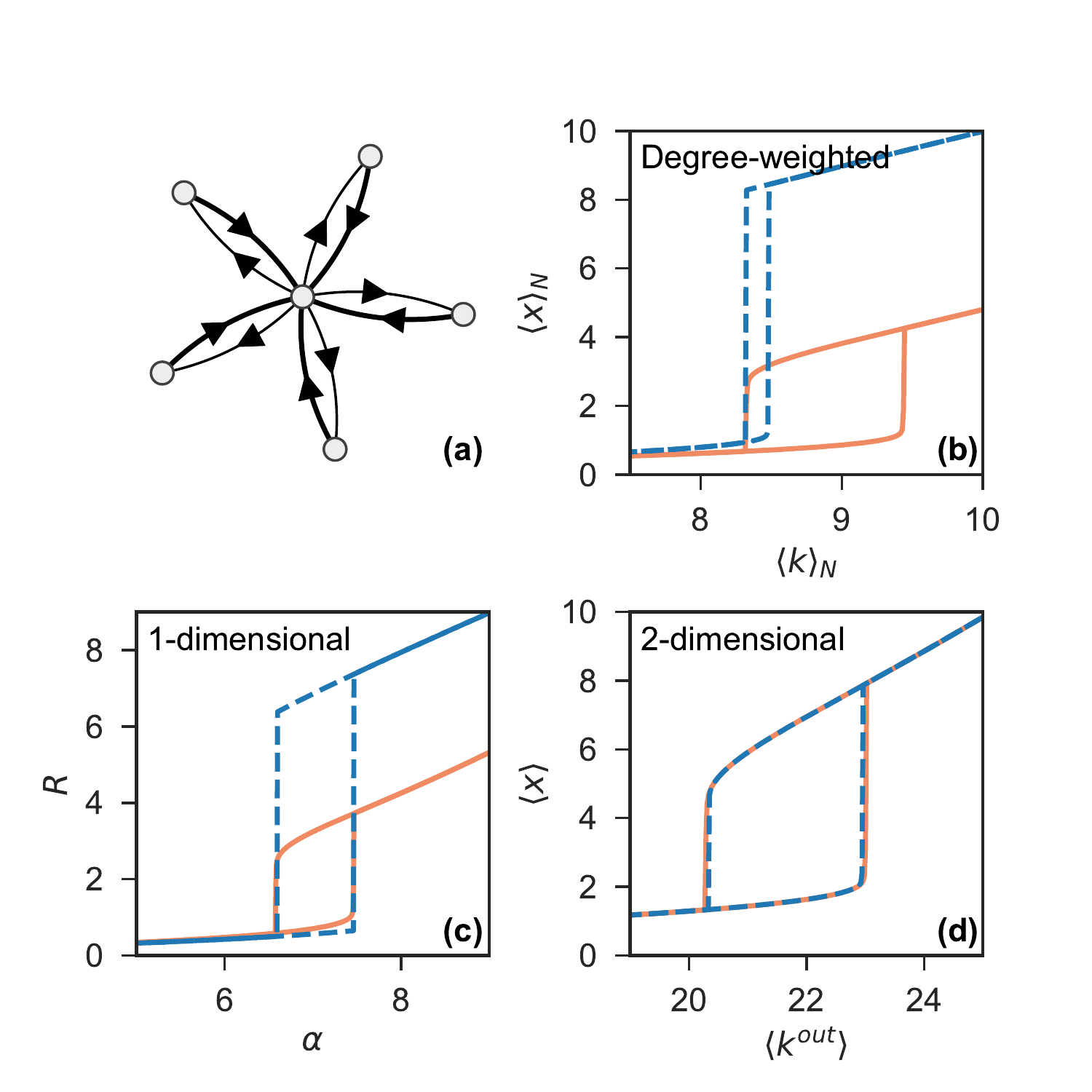}
  \caption{(Color online) (a) Schematisation of the star network of $N=6$ nodes where the edge weight toward the core is twice the weight of an edge toward the periphery. (b) Average neighbor activity at equilibrium as a function of its structural parameter using the degree-weighted reduction of Gao \textit{et al.}\cite{gao2016universal}. (c) Dominant eigenvector weighted activity at equilibrium $R^*=\bm{v}_D^T\bm{x}^*$ as a function of the dominant eigenvalue $\alpha$ for the 1--dimensional reduction [Eq.~\eqref {eq:R_dynamics}]. (d) Average activity at equilibrium obtained by a combination of two observables $\langle x\rangle^* = N^{-1}[R_1^*+(N-1)R_2^*]$, as a function of the average out-degree $\langle k^{\text{out}}\rangle=N^{-1}[\alpha_1+(N-1)\alpha_2]$ computed using the 2--dimensional reduction formalism [Eq.~\eqref{Eq:star_multi}] . Full lines are results from simulations and dashed lines are theoretical predictions. The network dynamics is the Cowan-Wilson model with $\tau=1,\mu=3$ [Eq.~\eqref{eq:cowan-wilson}].}
  \label{fig:star}
\end{figure}

%==============================================================================
\section{Multidimensional reduction: dominant eigenvectors}
\label{subsec:multidimensional}
%==============================================================================

In the 1--dimensional reduction, it has been supposed that the dynamical global state of a network is dominated by the information contained in a single dominant eigenvalue and corresponding eigenvector. Therefore, it also presupposes that other eigenvalues can be safely neglected and do not provide relevant information about the dynamics on the network. But, if the network admits many eigenvalues of similarly large modulus, it is plausible to expect that all these eigenvalues are important as well, and should be included in a $n$--dimensional reduction. In this section, we address this problem by introducing a $n$--dimensional approach to predict the evolution of $n$ coupled observables. We argue that if the spectrum $\{\lambda_1, \lambda_2, ..., \lambda_N\}$ of $\bm{W}^T$ satisfies $$|\lambda_1|\approx|\lambda_2|\approx...\approx|\lambda_n|\gg|\lambda_{n+1}|\geq...\geq|\lambda_{N}|,$$ then $n$ observables should be considered, leading to a $n$--dimensional reduced dynamical system.

%==============================================================================
\subsection{Cycle reduction}
\label{subsec:cycleapproach}
%==============================================================================

Let us consider $n$ observables $R_j$, $1\leq j\leq n$, each being a different linear combination of the activity
\begin{equation}
  R_j = \sum_{i=1}^{N}[\bm{a}_j]_i x_i,
\end{equation}
where $\bm{a}_j$ is a real-value weight vector associated with the observable $R_j$ and normalized $\sum_i [\bm{a}_j]_i=1$. As in the 1--dimensional reduction, $\bm{a}_j$ are yet undetermined. While there are several choices for $\bm{a}_j$ that are a priori plausible, we discuss the \textit{cycle} reduction, which is natural when several dominant eigenvalues are approximately of equal modulus.

Using a similar approach as the 1--dimensional reduction (See Appendix~\ref{sec:app:derivation-multidimensional}), one finds that the evolution of the observables is given by
\begin{equation}
  \dot{R}_j \approx \begin{cases}
F(R_j) + \alpha_j G(\beta_jR_j, R_{j+1}),~~~j<n\\
F(R_j) + \alpha_j G(\beta_jR_j, R_{1}),~~~~~~~j=n\\
\end{cases}
\label{eq:R_dimens}
\end{equation}
where 
\begin{equation}
  \beta_j = \dfrac{1}{\alpha_j}\dfrac{\bm{a}_j^T\bm{K}\bm{a}_j}{\bm{a}^T_j\bm{a}_j}\label{Eq:multi:beta}
\end{equation}
and $\alpha_j$ is an observable of the weighted neighbor in-degree,
\begin{equation}
  \alpha_j = \sum_{i=1}^N [\bm{a}_j]_i k^{\text{in}}_i.\label{eq:multi:alpha}
\end{equation}
To satisfy Eqs.~\eqref{eq:R_dimens}, the weight vectors are constrained by the structure and must transform according to
\begin{equation}
  \bm{a}_j = \dfrac{\bm{W}^T\bm{a}_{j-1}}{\alpha_{j-1}}\label{Eq:construction},
\end{equation}
preserves the positiveness and the required normalization. Moreover, Eq.~\eqref{Eq:construction} needs to be a periodic application, i.e. $\bm{a}_{j+n}=\bm{a}_j$, in order to close the system to $n$ observables. The initial choice of $\bm{a}_1$ is then highly constrained to satisfy this condition. In the following section, we will explain how the weight vectors can be computed using the dominant eigenvectors of $\bm{W}^T$.

In contrast to Eq.~\eqref{eq:R_dynamics} where a single observable is used, we now have developed a closed $n$--dimensional system of observables that are coupled by a set of structural parameters $\lbrace\alpha_j, \beta_j\rbrace_{j=1,..,n}$.

%==============================================================================
\subsection{Choice of the universal weight vectors}
\label{subsec:distinctobservables}
%==============================================================================

It is yet unclear if one should use a $n$--dimensional reduction or a 1--dimensional reduction for a certain network structure. By answering this question, we also address how to set the weight vectors of the cycle reduction.

Recall that the Perron-Frobenius theorem guarantees that $\bm{W}^T$  has a non-negative dominant eigenvector $\bm{v}_D$  only if $\bm{W}$ is a connected graph. Therefore, we can rule out that it is always possible to construct a 1--dimensional reduction relying on the dominant eigenvector.

But the same reasoning also implies that we can always construct a $n$--dimensional system using the dominant eigenvector $\bm{v}_D$ of $\bm{W}^T$. One could use $\bm{a}_1=\bm{v}_D$ and apply iteratively Eq.~\eqref{Eq:construction} to obtain the set of weight vectors $\lbrace \bm{a}_j\rbrace_{1, ...,n}$, as prescribed. In doing so, the resulting weight vectors would all be identical $\bm{a}_1=\bm{a}_2=...=\bm{a}_n=\bm{v}_D$, as it obviously satisfies both Eq.~\eqref{Eq:construction} and the periodicity condition $\bm{a}_{j+n}=\bm{a}_j$. Hence, we find $n$ identical observables $R_1=R_2=...=R_n$, and the constructed $n$--dimensional system is no better than the 1--dimensional system. For this reason, a $n$--dimensional cycle reduction is only advantageous if we can construct a set of \textit{distinct} weight vectors $\bm{a}_1\neq\bm{a}_2\neq...\neq\bm{a}_n$ from the dominant eigenvectors. 

The maximum number of significant and distinct observables that we can construct is determined by the periodicity of the transposed adjacency matrix $\bm{W}^T$. The periodicity $n$ is the number of eigenvalues $\lambda_m$ of modulus equal to the spectral radius $\spectralradius\geq 0$, i.e. $|\lambda_m|=\spectralradius$. From the Perron-Frobenius theorem, they must be uniformly distributed on a circle, centered at the origin, in the complex plane. Thus, the $m$\,th dominant eigenvalue can be written as
\begin{equation*}
	\lambda_m = \spectralradius\,\text{e}^{2\pi i m/n},
\end{equation*}
for a given periodicity $n$. Since $\lambda_m$ is an eigenvalue of $\bm{W}^T$, it must have an eigenvector $\bm{v}_m$ that satisfies
\begin{equation*}
	\bm{W}^T\bm{v}_m = \spectralradius\,\text{e}^{2\pi i m/n} \bm{v}_m.
\end{equation*}
By multiplying both sides by $(\bm{W}^T)^{(n-1)}$, we find
\begin{equation*}
	(\bm{W}^T)^n\bm{v}_m = \spectralradius^n \bm{v}_m.
\end{equation*}
Therefore, $\spectralradius^n$ is a real-value positive $n$ times degenerated eigenvalue of $(\bm{W}^T)^n$. Since each eigenvector of $\bm{W}^T$ is also eigenvector of $(\bm{W}^T)^n$, we can combine those eigenvectors to construct new distinct eigenvectors of $(\bm{W}^T)^n$ with eigenvalue $\spectralradius^n$ and use them as weight vectors. We construct the first weight vector as
\begin{equation}
  	\bm{a}_1 = \dfrac{\sum_{m=1}^n c_{m}\bm{v}_m}{\sum_{m=1}^nc_m\bm{1}^T\bm{v}_m},
\end{equation}  
where $c_{m}\in\mathbb{C}$ are arbitrary coefficients. From Eq.~\eqref{Eq:construction}, we iteratively compute $\bm{a}_j$ from $\bm{a}_{j-1}$. 
By doing so, we both satisfy the periodic condition $\bm{a}_{j+n}=\bm{a}_j$ and construct distinct weight vectors.

The reduction only requires to arbitrarily choose $\mathbf{c}=(c_1,\ldots,c_n)$ to construct $\bm{a}_1$. We propose to select $\bm{c}$ by minimizing the scalar product of the first two weight vectors
\begin{align}
	\bm{c} = \underset{\bm{c}}{\mathrm{argmin}}|\bm{a}_1^T\bm{a}_2|\label{Eq:LSE}
\end{align}
where $\bm{a}_2=\alpha_1^{-1}\sum_{m=1}^n c_{m}\lambda_m \bm{v}_m$ from Eq.~\eqref{Eq:construction}. In Appendix~\ref{sec:app:analytics-solution}, we give a general and exact solution of $\bm{c}$ for $n=2$.

In summary, the cycle reduction method goes as follows. First, compute a set of $n$ eigenvectors $\lbrace \bm{v}_m\rbrace_{m=1,...,n}$ of $\bm{W}^T$ whose eigenvalues have a modulus equal to the spectral radius $\spectralradius$. Second, obtain $\lbrace c_{i}\rbrace_{i=1,...,n}$ by solving Eq.~\eqref{Eq:LSE}. Third, iteratively construct $\bm{a}_i$ from Eq.~\eqref{Eq:construction}. Finally, compute $\alpha_i, \beta_i$ and solve $R_i$ at equilibrium from Eqs.~\eqref{eq:R_dimens}. 

An interesting aspect of this method is that it allows to combine the information of each observable to construct a global observable
\begin{equation}
	R_{\text{global}}=\sum_{j=1}^n \phi_jR_j,
\end{equation}
where $\phi_j\in\mathbb{R}$. Since $R_j=\sum_i [\bm{a}_i]_j x_j$,    the contribution $p_j$ of node $j$ to the global observable is 
\begin{equation}
	p_j=\sum_i \phi_i[\bm{a}_i]_j.
\end{equation}
We can then tune $\phi_i$ to reach the desired node contributions. For instance, to access the unweighted average activity $R_{\text{global}}=\langle x\rangle$, one solves $\bm{A}\bm{\phi}=N^{-1}\bm{1}_N$, where $\bm{A}=[\bm{a}_1~\bm{a}_2~...~\bm{a}_n]$. The solution is $\bm{\phi}=N^{-1}\bm{A}^+\bm{1}_N$, where $\bm{A}^+$ is the Moore-Penrose pseudo-inverse of matrix $\bm{A}$ \cite{horn1990matrix}.

%==============================================================================
\subsection{Examples: Star and bipartite networks}
\label{subsec:starnetworks}
%==============================================================================

We give an example of the cycle reduction for a highly heterogeneous family of networks: star networks. We construct a star network of $N$ nodes where the strength of a directed edge to a periphery nodes is $\toperi$ and $\tocore$ for edges directed toward the central node. Hence, the adjacency matrix is
\begin{equation*}
 \bm{W} = \begin{bmatrix}
    0 & \tocore & \tocore & \dots  & \tocore \\
    \toperi & 0 & 0 & \dots  & 0 \\
    \vdots & \vdots & \vdots & \ddots & \vdots \\
    \toperi & 0 & 0 & \dots  & 0
\end{bmatrix}.
\end{equation*}
One finds that $\bm{W}^T$ has two eigenvalues of modulus equal to the spectral radius, $\lambda_+=\sqrt{\toperi\tocore(N-1)}$ and $\lambda_-=-\sqrt{\toperi\tocore(N-1)}$. From the previous analysis, this signals that we can construct a $2$--dimensional reduction. The associated eigenvectors are
\begin{subequations}
\begin{align}
	\bm{v}_+^T &=\left[\dfrac{\lambda_+}{\tocore}~~1~~1~~...~~~1  \right],\\
	\bm{v}_-^T &=\left[\dfrac{\lambda_-}{\tocore}~~1~~1~~...~~~1   \right].
\end{align}
\end{subequations}
We construct the $2$--dimensional reduction by combining the eigenvectors $\bm{v}_+, \bm{v}_-$ to minimize $|\bm{a}_1^T\bm{a}_2|$ and to satisfy the normalization $\bm{1}^T\bm{a}_1=1$. The following linear combinations fulfill the requirements:
\begin{align}
	\bm{a}_1 &= \dfrac{\bm{v}_+-\bm{v}_-}{\bm{1}^T\bm{v}_+-\bm{1}^T\bm{v}_-},\\
	\bm{a}_2&=\dfrac{1}{(N-1)\tocore}\dfrac{(\lambda_+\bm{v}_+-\lambda_-\bm{v}_-)}{\bm{1}^T\bm{v}_+-\bm{1}^T\bm{v}_-},
\end{align}
where the second vector has been obtained from Eq.~\eqref{Eq:construction}. Note that the overlap $\bm{a}_1^T\bm{a}_2=0$ exactly. Explicitly, in component form, we have
\begin{subequations}
\begin{align}
  [\bm{a}_1]_i &= \delta_{i,1},\\
  [\bm{a}_2]_i &= \dfrac{1}{N-1}(1-\delta_{i,1}).
  \label{Eq:star:group}
\end{align}
\end{subequations}
One also finds that $\beta_1=1,\beta_2=1, \alpha_1=\tocore(N-1), \alpha_2=\toperi$.
Hence, the 2--dimensional reduction reads
\begin{subequations}
\begin{align}
  \dot{R}_1 &= F(R_1)+\tocore(N-1)G(R_1, R_2),\\
  \dot{R}_2 &= F(R_2)+\toperi G(R_2, R_1).\label{Eq:star_multi}
\end{align} 
\end{subequations}
One notes that $R_1$ is exactly equal to the activity of the central node and $R_2$ is the activity of a periphery node. Thus, the 2--dimensional formalism is an exact reduction in this example. This is confirmed in Fig.~\ref{fig:star} where simulations and predictions are compared for the Gao \textit{et al.}, the 1--dimensional reduction and the 2--dimensional reduction for the Cowan-Wilson dynamics.

\begin{figure}[t!]
  \centering
  \includegraphics[width=\linewidth]{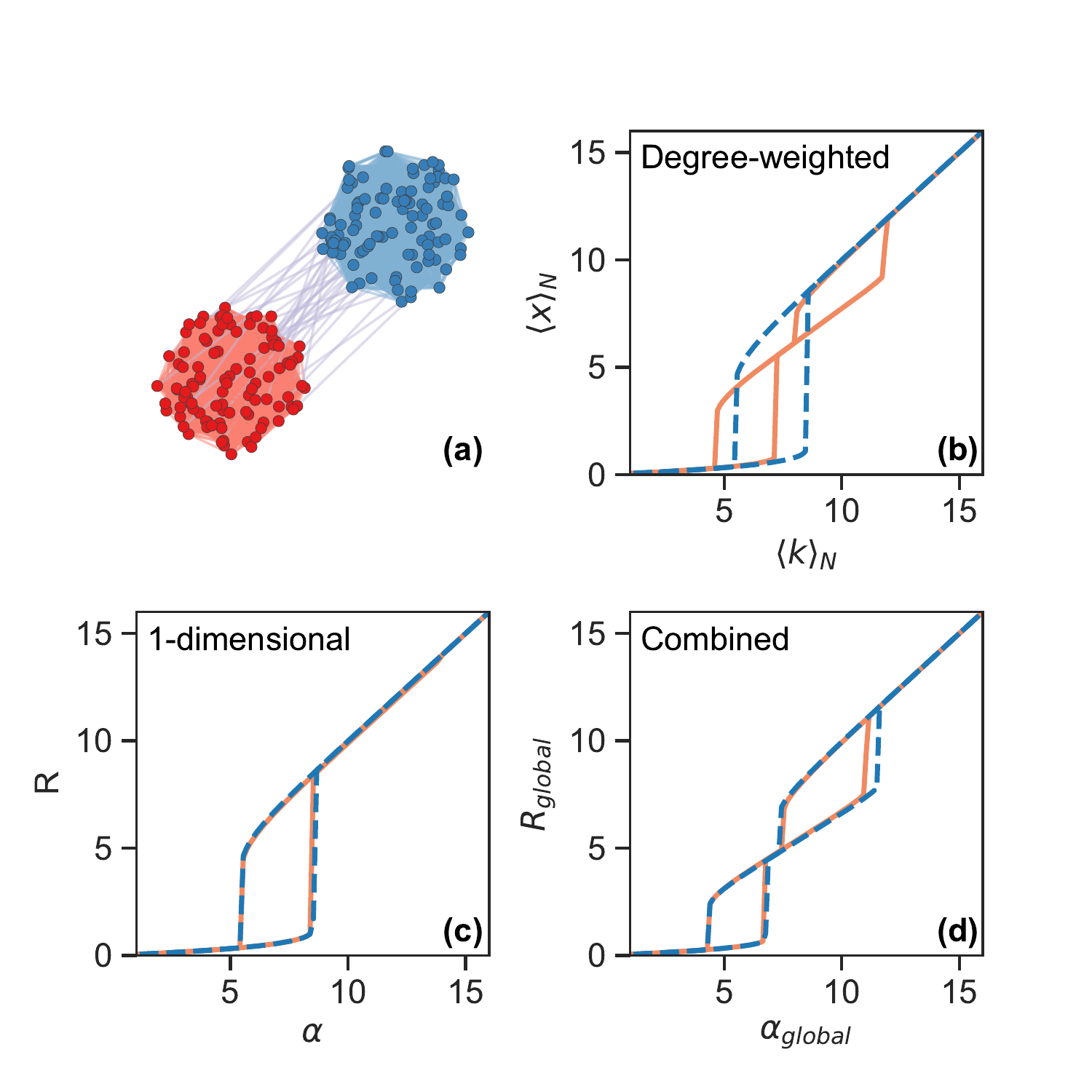}
  \caption{(Color online) (a) Schematisation of the undirected planted partition network of two communities of 100 nodes each with in-densities $p_{\text{in}}=0.4$ and $p_{\text{in}}=0.7$, and out-density $p_{\text{out}}=3\times10^{-3}$. (b) Average neighbor activity at equilibrium as a function of its structural parameter using the degree-weighted reduction of Gao \textit{et al.} \cite{gao2016universal}. (c) Dominant eigenvector-weighted activity $R=\bm{v}_D^T\bm{x}$ at equilibrium as a function of the associated eigenvalue $\alpha$ for the 1--dimensional reduction [Eq.~\eqref{eq:uni:equilibirum}]. (d) Combination of the uncoupled observables $R_{\text{global}}=(R_D+R_{SD})/2$ at equilibrium as a function of the structural parameter $\alpha_{\text{global}} $ [See Eq.~\eqref{Eq:sbm_uncoupled}].  Full lines result from simulations and dashed lines are theoretical predictions. The dynamics is the Cowan-Wilson model with $\tau=1,\mu=3$ [Eq.~\ref{eq:cowan-wilson}].}
  \label{fig:sbm}
\end{figure}

Star networks are not the only systems conforming to the 2--dimensional reduction; all bipartite networks also do. Bipartite networks have nodes that can be separated into two groups such that connections only exist between nodes of different groups \cite{newman2018networks}. This network architecture is common in many real systems such as plant-pollinator interactions \cite{bastolla2009architecture}, scientific collaborations \cite{newman2001scientific}, and actor-film networks \cite{watts1998collective}.

Bipartite networks exhibit a remarkable and useful spectral property. Each eigenvalue is paired, i.e. 
\begin{equation}
	\lambda_j=-\lambda_{N-j+1},
\end{equation}
for all $j=1,2,..,N$, assuming that $\lambda_1\geq\lambda_2\geq...\geq\lambda_N$ \cite{cvetkovic1980spectra}. Thus, a bipartite graph contains two eigenvalues $\lambda_1, \lambda_N$ of modulus equal to the spectral radius. This suggests that a 2--dimensional representation could always be constructed out of these two eigenvalues, following the prescription of Sec.~\ref{subsec:cycleapproach}.

We therefore gain a clear understanding of the results of Jiang \textit{et al.}\cite{jiang2018predicting}, where they present numerical evidences on real bipartite networks, suggesting that 2--dimensional reductions are better predictors of critical points for bipartite networks than the 1--dimensional approaches. Moreover, it solves the problem of selecting the right weighted combination: The eigenvector-weighted combination
 should always be favored over others.

\begin{figure*}[t!]
  \centering
  \includegraphics[width=0.94\linewidth]{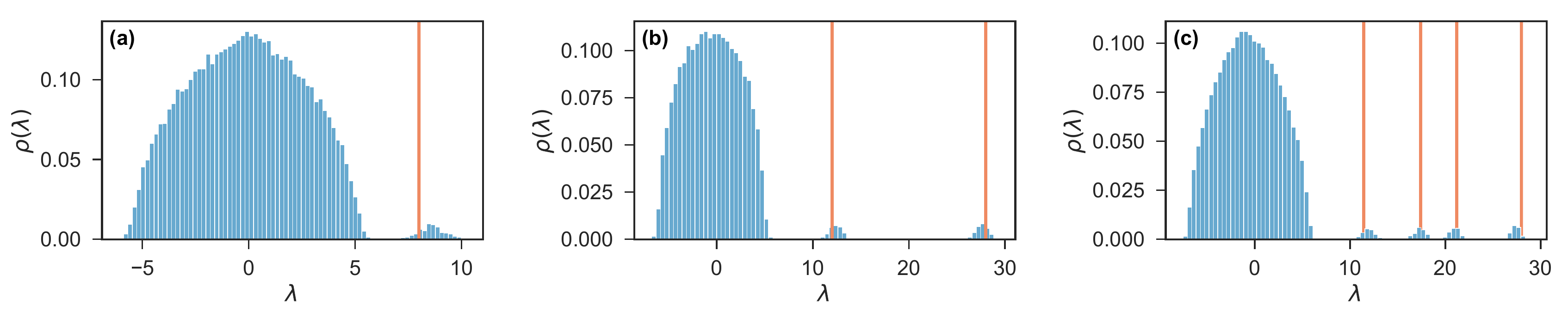}
  \caption{(Color online) Spectral density $\rho(\lambda)$ for (a) random networks of $80$ nodes with connection probability $p=0.1$, 
(b) of SBM with two communities of 40 nodes with in-connection probabilities $p_{11}=0.3$ and $p_{22}=0.7$ and out-connection probabilities $p_{12}=p_{21}=0.01$, and 
(c) of SBM with four communities of 40 nodes with in-connection probabilities $p_{11}=0.4$, $p_{22}=0.6$, $p_{33}=0.7$,  and $p_{44}=0.9$ and out-connection probability $p_{rs}=0.05~\forall r\neq s$. The orange lines highlight the position of the dominant and subdominant eigenvalues in the expected network. Each spectral density is produced by collecting spectra from 500 network instances. }
  \label{fig:planted-eigenvalues} 
\end{figure*}

%==============================================================================
\section{Multidimensional reduction: including subdominant eigenvectors}
\label{sec:linear combinaison}
%==============================================================================

Until now, we have developed a direct method to construct $n$--dimensional reduced systems. Using only the network structure, we can first identify the number of dimensions, i.e. the number of eigenvalues of modulus equal to the spectral radius, and then construct the weight vectors $\bm{a}_j$ to predict the observables $R_j$. 

Yet, our method is compelling only if the observables $R_j$ are good indicators of the global states of the network, which requires that each region of the network contributes significantly to at least one observable $R_j$. Since we use the dominant eigenvectors, we do not control the contribution of each node. Thus, if the dominant eigenvector assigns negligible weights to some nodes, it may result in an incomplete description of the network. Modular networks fall into this category. Let us introduce the stochastic block model (SBM) to understand the underlying problem of misrepresentation.

The SBM is a generative model of modular networks \cite{holland1983stochastic}. Nodes are first assigned to modules. Then, we connect a node from module $s$ to a node from module $r$ with probability $p_{rs}$. This simple method generates accurately modular random networks.

The spectrum of a SBM network is rather different than the spectrum of a random network (Fig.~\ref{fig:planted-eigenvalues}). We first note that we only have a single eigenvalue of modulus equal to the spectral radius, indicating that we should use a 1--dimensional reduction. However, the eigenvalues are distributed in a multimodal distribution with as many dominant and subdominant eigenvalues as they are modules. 

For instance, let us consider a network with two communities of equal size $N/2$. Using the spectral theory of random matrices, we estimate the two dominant eigenvalues
\begin{align*}
	\lambda_1 &= \dfrac{(p_{11}+p_{22})+[(p_{11}-p_{22})^2 + 4p_{12}p_{21}]^{1/2}}{2},\\
	\lambda_2 &= \dfrac{(p_{11}+p_{22})-[(p_{11}-p_{22})^2 + 4p_{12}p_{21}]^{1/2}}{2},
\end{align*}
and their corresponding eigenvectors
\begin{align*}
	\bm{v}_1 = \begin{bmatrix}\dfrac{p_{12}}{p_{12}+\lambda_1-p_{11}}\bm{1}_{N/2}\\\\ \dfrac{\lambda_1-p_{11}}{p_{12}+\lambda_1-p_{11}}\bm{1}_{N/2}\end{bmatrix},
\end{align*}

\begin{align*}
	\bm{v}_2 = \begin{bmatrix}\dfrac{\lambda_2-p_{22}}{\lambda_2-p_{22}+p_{21}}\bm{1}_{N/2},\\\\ \dfrac{p_{21}}{\lambda_2-p_{22}+p_{21}}\bm{1}_{N/2}\end{bmatrix}.
\end{align*}
If $p_{rs}/p_{rr}\approx 0~~\forall r \neq s$, the dominant eigenvector $\bm{v}_1$ assigns a negligible weight to the nodes in the community $r=1$~\footnote{Note that if $(p_{11}-p_{22})^2\ll 4p_{12}p_{21}$, then the problem of misrepresentation is absent and it reduces to a single 1--dimensional representation.}.Thus, if we solely use a 1--dimensional reduction with $\bm{a}=\bm{v}_1$, the observable $R$ will not take into account the activity of half the network. Fortunately, the second-dominant eigenvector accounts for the remaining nodes. Thus, we must apply the 1--dimensional reduction of Sec.~\ref{subsec:1--dimensional} twice and construct two uncoupled observables, one with the dominant eigenvector $R_{D}=\bm{v}_1^T\bm{x}$, and one for the subdominant eigenvector $R_{SD}=\bm{v}_2^T\bm{x}$, for which the dynamics follow,
\begin{subequations}
\begin{align} 
	\dot{R}_D&=F(R_D)+\alpha_D G(\beta_D R_D, R_D),\\
	\dot{R}_{SD}&=F(R_{SD})+\alpha_{SD}G(\beta_{SD}R_{SD}, R_{SD}).\label{Eq:sbm_uncoupled}
\end{align}
\end{subequations}
To make a global prediction, we simply combine the observables
\begin{equation}
	R_{\text{global}} = \dfrac{R_D+R_{SD}}{2}.\label{Eq:combiningobservable}
\end{equation}
It follows that the global structural parameter is also a linear composition
\begin{equation}
	\alpha_{\text{global}}  = \dfrac{\alpha_D+\alpha_{SD}}{2}.
\end{equation}
In general, we can construct as many uncoupled observables as the number of modules in the network, by using the eigenvectors associated with the eigenvalues detached from the bulk of the spectrum.

\begin{figure*}[t!]
	\centering
	\includegraphics[width=\linewidth]{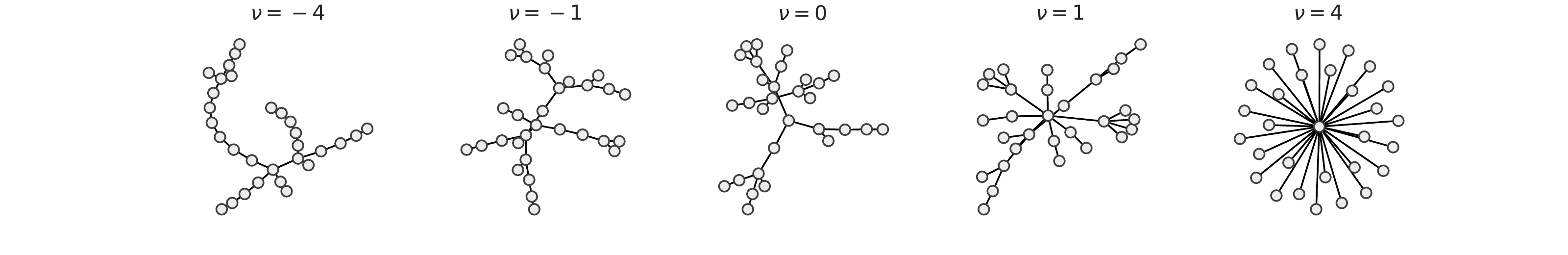}
	\caption{(Color online) Network instances produced using the generalized preferential attachment model. The model tends to generate chain-like networks for negative values of $\nu$ and star networks for positive values. Figure inspired from \cite{young2018network}.}
	\label{fig:generalized}
\end{figure*}

We show a numerical example of this method for a network of two communities (Fig~\ref{fig:sbm}). The Gao \textit{et al.} formalism predicts a single bifurcation, which almost coincides with the transition of the densest community. In Fig~\ref{fig:sbm}(b), the 1--dimensional reduction, using the first dominant eigenvector, predicts the activity accurately. However, the first dominant eigenvector omits half the network and we only see a single bifurcation. Thus, even if we are highly accurate, we miss characterizing the distinctive multistep bifurcation, a prominent feature of interacting networks \cite{majdandzic2016multiple}. Finally, combining observables as in Eq.~\eqref{Eq:combiningobservable}, we recover the bifurcations of both modules [Fig~\ref{fig:sbm}(d)].

%==============================================================================
\section{Goodness of reduction}
\label{subsec:scalefree}
%==============================================================================

The goodness of the reduction method, i.e. how accurate are the predictions of the low-dimensional representation compared with the observations on the original network, depends on the nature of the dynamics and on the network structure. For instance, some complex patterns of interactions may be less amenable to a low-dimensional formalism, resulting in disparities between the predicted and exact values of the observables. In this section, we explore the impacts of the structure and the dynamics on the goodness of the reduction methods.

%==============================================================================
\subsection{Impact of the structure}
\label{subsec:scalefree:avrerror}
%==============================================================================

To measure the impact of the structure, we introduce a generative model of networks called \textit{generalized preferential attachment model} \cite{krapivsky2000connectivity}. Parameters of this model can be continuously tuned to obtain networks ranging from chain-like networks to star networks, with scale-free networks as an intermediate state. Scale-free systems are an important family of networks, recognizable by their power-law degree distribution $p(k)\sim k^{-\gamma}$ \cite{barabasi1999emergence}. Due to their lack of well-defined characteristic scale, it is a priori unclear whether these systems can be efficiently reduced. 

The growth process of the generalized preferential attachment goes as follows. We initialize the network with two connected nodes. Then, at each time step $t$, we add a new node to the network. It is connected to an existing node chosen with probability
\begin{equation}
	w_i(\nu, t)=\dfrac{s_i^{\nu}(t)}{\sum_{j=1}^{N(t)}s_j^{\nu}(t)}\label{Eq:scalefreeprob}
\end{equation}
where $\nu\in\mathbb{R}$ is the exponent of the attachment kernel, $s_j(t)$ is the number of connections of node $j$ at time $t$, and $N(t)$ is the number of nodes at time $t$.

The generative model is solely tuned by the kernel parameter $\nu\in\mathbb{R}$. It controls the inequalities of the attachment probability. On the one hand, if $\nu\gg1$, the generated networks are star-like as we always attach new nodes to the richest node. On the other hand, if $\nu\ll0$, the networks are more chain-like as we always connect to the least connected node \cite{krapivsky2000connectivity}. The classic preferential attachment model is found for $\nu=1$. Therefore, for $0<\nu <1$, we observe a continuum of network organizations which gradually become more scale-free as the parameter $\nu$ is increased. Examples of networks generated from this model are illustrated in Fig.~\ref{fig:generalized}.

From now on, we will distinguish the predicted observable from the reduced system, denoted $\tilde{R}(\alpha)$, and the measured observable $R_{\obs}(\alpha)=\bm{a}^T\bm{x}$ from the original network.

We have applied the degree-weighted, the 1--dimensional and the 2--cycle reductions to networks generated with the generalized preferential attachment model for $\nu\in[-1,2]$. For each network, we have computed the total error $\Delta_R$ between the measured activity $R^*_{\obs}(\alpha)=\bm{a}^T\bm{x}^*$ at equilibrium on the original network and the predicted activity $\tilde{R}^*(\alpha)$ by the reduction system,
\begin{equation}
	\Delta_R = \int_{0}^{\infty}|\tilde{R}^*(\alpha)-R^*_{\obs}(\alpha)|d\alpha\label{Eq:avrerror}.
\end{equation}  

\begin{figure}[t!] 
  \centering
  \includegraphics[width=0.95\linewidth]{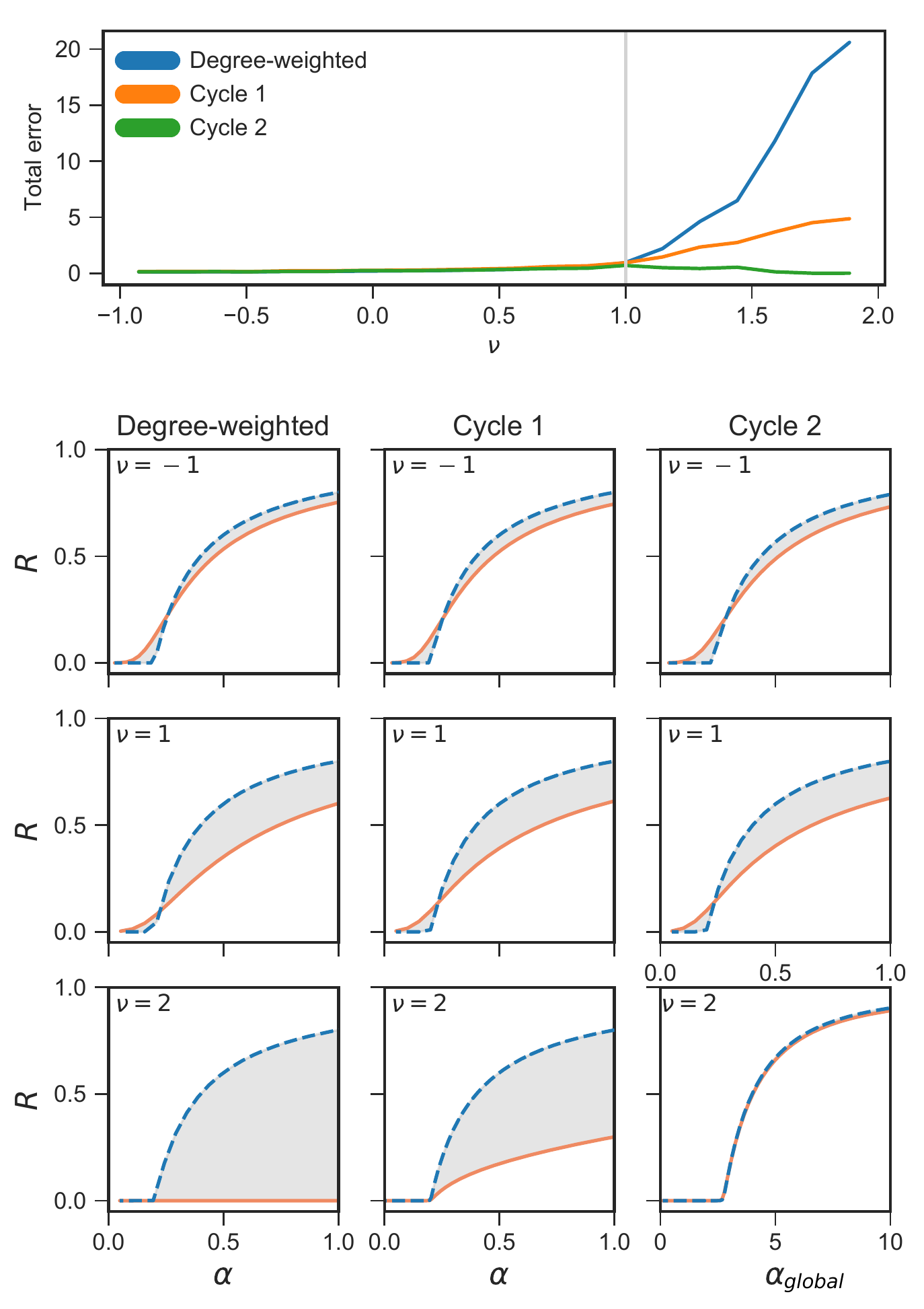}
  \caption{(Color online) (Top) Total error on the prediction of the global state activity at equilibrium for networks obtained from the generalized preferential attachment model [See Eq.~\eqref{Eq:avrerror}]. A gray line indicates the classical preferential attachment model at $\nu=1$ where a critical transition in the reduced descriptions is found. The total error is averaged from 300 networks of $N=200$ generated uniformly on the domain $\nu\in[-1,2]$. The activity on the network is the SIS model with $\gamma=1$ [See Eq.~\eqref{eq:sis}]. (Bottom) Instances of bifurcation diagrams for the three reduction schemes (columns) and for $\nu=\lbrace -1,1,2\rbrace$ (rows). For the 2--dimensional cycle reduction, the x-axis is the average of the structural parameters: $\alpha_{global}=(\alpha_1+\alpha_2)/2$. The blue dashed lines are predictions from reduced systems and the orange lines are the measured activities on the original networks. The gray regions indicate the absolute errors [See Eq.~\eqref{Eq:avrerror}].}
  \label{fig:pa-error}
\end{figure}

We have found a transition in the dimension reduction accuracy for all methods at $\nu=1$, corresponding to the preferential attachment model [Fig~\ref{fig:pa-error}]. As we enter the star-like region $\nu>1$, the average error reaches a plateau to specific values for the $1$--dimensional reductions while the 2--dimension reduction remains highly effective.

We argue that this transition is not dynamics-specific but mostly due to the network architecture. The nature of the error can be interpreted by a careful examination of the generative model. First, negative values of $\nu$ tend to homogenize the degree of the nodes. Thus, the more uniform the network is, the easier it is to capture its behavior in a 1--dimension reduction. For positive values of $\nu$, the reduced model tends to favor degree inequalities which are best achieved when the networks are star-like.

For $0<\nu<1$, the degree distribution resembles a power law with exponential cutoffs \cite{krapivsky2000connectivity}. However, a pure power-law distribution is only achieved precisely at $\nu=1$. Thus, this transition in the degree distribution forces the reduction of the degree-weighted reduction to predict inaccurate observables. Finally, the region $\nu>1$ is dominated by star-like networks, which has been previously shown to be better represented by the 2--dimensional reduction than any 1--dimensional reduction (See Subsec.~\ref{subsec:starnetworks}).

%==============================================================================
\subsection{Detection of transitions}
\label{subsec:scalefree:avrerror2}
%==============================================================================
Most dynamical systems exhibit activity bifurcations when a certain structural threshold is reached \cite{weisbuch2018complex}. Therefore, the goodness of the reduction should display at least qualitative changes as the structural threshold is crossed. We investigate this kind of prediction using the SIS model.

The SIS model has been thoroughly studied in the last decade \cite{pastor2015epidemic}. In the SIS model, nodes reversibly switch from susceptible to infected states with a certain probability that depends on their neighborhood. We can formulate this dynamics using a mean-field approach,
\begin{equation}
	\dot{x}_i=-x_i+\gamma(1-x_i)\sum_{j=1}^Nw_{ij}x_j,
\end{equation}
where $x_i$ is the probability that node $i$ is infected and $\gamma\geq 0$ is the normalized infection rate. In this model, the average fraction of infected node $\langle x\rangle=N^{-1}\sum_i x_i$ undergoes a critical transition at a certain threshold $\gamma_C$. The classical problem in the study of the SIS model consists in estimating the value of $\gamma_C$ above which a significant fraction of the whole system is infected \cite{st2018phase}.  

\begin{figure}[t!]
	\centering
	\includegraphics[width=\linewidth]{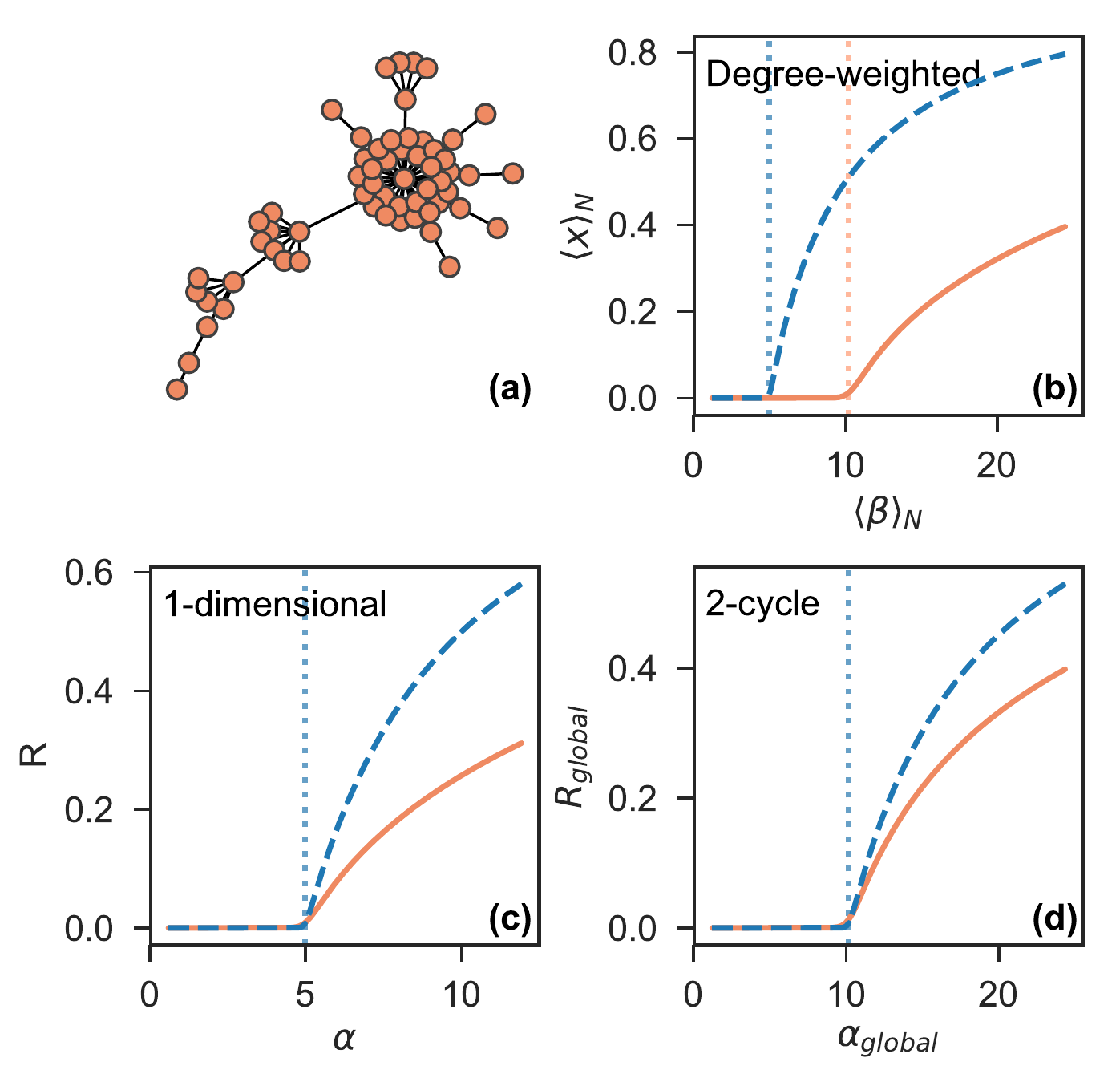}
	\caption{(Color online) (a) Schematisation of the undirected network generated using the generalized preferential attachment model with $\nu=1.8$ and $N=60$. (b) Average neighbor activity at equilibrium as a function of its structural parameter using the degree-weighted reduction \cite{gao2016universal}.  (c) Dominant eigenvector-weighted activity at equilibrium $R^*=\bm{v}_D^T\bm{x}^*$ as a function of the associated eigenvalue $\alpha$ for the 1--dimensional reduction [Eq.~\eqref {eq:R_dynamics}]. (d) Solution at equilibrium of the 2--dimensional system $R_{\text{global}}^*=(R^*_1+R^*_2)/2$ as a function of the structural parameter $\alpha_{\text{global}}=(\alpha_1+\alpha_2)/2$. Full lines are results from simulations and dashed lines are theoretical predictions. Dotted lines indicate the position of the transition $\alpha_{\text{global}} $. The dynamics is the SIS model with $\gamma=0.2$ [Eq.~\eqref{eq:sis}]. }
	\label{fig:sis}
\end{figure}

We will however study a related problem: the parameter $\gamma$ is fixed and the structure is evolving. Using the dimension reduction procedure, we investigate the critical structural parameter $\alpha_{\text{global}}$, or an equivalent parameter depending on the reduction approach, for which the global state at equilibrium $R^*$ undergoes a critical transition characterized by
\begin{equation}
	\dfrac{d^2R^*(\alpha_C)}{d \alpha^2}=0.
\end{equation}
In Fig.~\ref{fig:sis}, we investigate the errors on the position of the critical transition for the degree weighted 1--dimension approach, the eigenvector-weighted 1--dimensional approach, and the 2--dimensional \textit{cycle} reduction. We use a network of $N=60$ nodes generated from the generalized preferential attachment model with $\nu=1.8$. We observe that the two proposed approaches based on dominant eigenvectors are able to accurately predict the critical transition while the degree-weighted approach does not. This behavior is typical for reductions of networks generated from $\nu\in[1,\infty)$. Our results indicate that even if the 1--dimensional observable fails to predict the true level of activity $R$, it still predicts with high accuracy the onset of the epidemy. We conclude that the largest eigenvalue is a reliable indicator of the onset for correlated networks. Perhaps, this conclusion is not surprising as it has been previously discovered under a different approach \cite{boguna2002epidemic}. Nonetheless, it supports the proposed reductions as valuable candidates for predicting the onset of critical transitions.

%==============================================================================
\subsection{Impact of the dynamics}
\label{sec:error}
%==============================================================================

As previously discussed, the goodness of the reduction is highly dependent on the network structure. The other element that impacts the goodness of the reduction is the nature of the dynamics. For instance, linear dynamics such as $F(x_i)=x_i$, $G(x_i,x_j)=x_j$ lead to exact 1--dimensional reduction. However, typical dynamics are nonlinear and may add significant contributions to the quadratic terms that have been neglected (See Appendix \ref{sec:app:derivation}). We investigate this aspect by looking at two contrasting dynamics: Cowan-Wilson model and Lotka-Volterra model. 
 
First, let us introduce the relative error $\Delta_{\alpha}$ on the structural parameter that predicts the original network activity. We compare the measured structural parameter $\alpha_{\obs}=\bm{a}^T\bm{k}^{\text{in}}$ on the original network with the structural parameter $\tilde{\alpha}$ that matches, from the 1--dimensional reduction, the measured activity $R^*_{\obs}=\bm{a}^T\bm{x}^*$ on the original network. The relative error can be written as
\begin{equation*}
	\Delta_{\alpha} = \dfrac{\alpha_{\obs}-\tilde{\alpha}(R^*_{\obs} )}{\alpha_{\obs }}.
\end{equation*}
Notice that this is different from Eq.~\eqref{Eq:avrerror}: $\Delta_{\alpha}$ is the horizontal error in the space $(\alpha, R)$ while $\Delta_R$ measures the vertical error. Both errors convey distinct information and are complementary. However, we will use $\Delta_{\alpha}$ since it can be written as a function that depends explicitly on the nature of the dynamics for the $1$--dimensional reduction. 

From Eq.~\eqref{eq:R_dynamics}, we obtain $\tilde{\alpha}$ at the dynamical equilibrium $\dot{R}=0$. This leads to
\begin{equation}
	\Delta_{\alpha} = 1+\dfrac{F(R^*_{\obs })}{G(R^*_{\obs })\alpha_{\obs }}\label{eq:error1}.
\end{equation}
One can then evaluate the error for specific dynamics as in \cite{tu2017collapse}. In the following paragraphs, we give two examples of dynamics and compare the errors for the Gao \textit{et al.} formalism and our 1--dimensional reduction.

%==============================================================================
\subsubsection{Error on the Cowan-Wilson dynamics}
\label{subsec:error:CW}
%==============================================================================

The Cowan-Wilson dynamics describes the firing-rate activity of populations of neurons. The evolution of a node activity is given by Eq.~\eqref{eq:cowan-wilson}, and repeated here as
\begin{equation}
	\dot{x_i}=-x_i+\sum_{j=1}^Nw_{ij}\dfrac{1}{1+\exp[-\tau(x_j-\mu)]},
\end{equation} 
where $\tau>0$. The equilibrium solution $\bm{x}^*$ cannot be found analytically, so it must be evaluated numerically, even for $N=1$. However, $\bm{x}^*$ is well approximated for extreme values of activities. We derive error estimates for extreme regimes $x_j^*\gg\mu$ and $x_j^*\ll \mu$.

In general, from Eq.~\eqref{eq:error1}, the error can be written as
\begin{equation}
	\Delta_{\alpha} = 1-\dfrac{[1+\exp[-\tau(R^*_{\obs }-\mu)]]R^*_{\obs }}{\alpha_{\obs }}.
\end{equation}
In the limit of high levels of activity $x_j^*\gg\mu$, the exponential vanishes and one finds that $\bm{x}^*\approx \bm{W}\bm{1}=\bm{k}^{\text{in}}$. Therefore, the error is approximately
\begin{align*}
	\Delta_{\alpha}\approx 1-\dfrac{R^*_{\obs }}{\alpha_{\obs }}\approx 1 - \dfrac{\bm{a}^T\bm{W}\bm{1}}{\alpha_{\obs }}.
\end{align*}
For the 1--dimensional reduction, $\bm{a}$ is an eigenvector of $\bm{W}^T$ such that $\bm{a}^T\bm{W}=\alpha\bm{a}^T$. Since $\bm{a}$ is normalized, $\bm{a}^T\bm{1}=1$, it follows that $\bm{a}^T\bm{W}\bm{1}=\alpha_{\obs}$. Thus, the error $\Delta_{\alpha}$ vanishes in the limit of large activity.

The same applies for Gao \textit{et al.} formalism for high activity. By using $\bm{a}=\bm{k}^{\text{in}}$, one finds that $\alpha_{\obs}=(\bm{k}^{\text{in}})^T\bm{k}^{\text{out}}$ and $R_{\obs}^*=(\bm{k}^{\text{out}})^T\bm{k}^{\text{in}}$ so that $\Delta_{\alpha}\to0$. 

Using a similar procedure for $x_j\ll \mu$, one can also show that $\Delta_{\alpha}\approx 0$ for both methods.

We conclude that in the extreme regimes of high and low activities, both reduction methods provide a practically exact solution. However, we are more often interested in the hysteresis region where the activity collapses rapidly. Unfortunately, analytic error estimates are lacking in this regime. Still, numerical results and theoretical insights suggest that the proposed 1--dimensional reduction should always be favored over the degree-weighted reduction.
This is confirmed below for a more tractable dynamics.

\subsubsection{Error on the Lotka-Volterra dynamics}
Let us consider the Lotka-Volterra dynamics governing the evolution of species populations. The $N$--dimensional system goes as 
\begin{equation}
	\dot{\bm{x}}= \omega \bm{x}+\bm{x}\circ \bm{W}\bm{x},
\end{equation}
where $\circ$ denotes an elementwise multiplication.
At equilibrium, $\bm{x}^*$ satisfies
\begin{equation}
	-\omega\bm{1} = \bm{W}\bm{x}^*.\label{eq:lokta:solution}
\end{equation}
With Eq.~\eqref{eq:error1}, we write the expected error as
\begin{align*}
	\Delta_{\alpha}  = 1 + \dfrac{\omega}{\beta \alpha_\obs R^*_{\obs }}=1 + \dfrac{\omega}{\beta \alpha_\obs \bm{a}^T\bm{x}^*}.
\end{align*}
Using the 1--dimensional reduction, 
\begin{equation*}
	\bm{W}^T\bm{a}=\alpha_{\obs }\bm{a},
\end{equation*}
or $\bm{a}^T\bm{W}=\alpha_{\obs }\bm{a}^T$. Furthermore, 
\begin{equation}
	\bm{a}^T\bm{W}\bm{x}^* = -\omega\bm{a}^T\bm{1}= \alpha_{\obs }\bm{a}^T\bm{x}^* .
\end{equation}
Thus, the error depends only on $\beta$:
\begin{equation}
	\Delta_{\alpha} = 1 - \dfrac{1}{\beta}.
\end{equation}
Therefore, the difference between the exact value for $\bm{a}^T\bm{x}^*$ and the approximate value derived from the 1--dimensional reduced system is only $1-1/\beta$. Given the expression for $\beta$ [Eq.~\eqref{eq:beta:ratio}], it should be close to $\beta\approx1$, so the error goes to zero $\Delta_{\alpha}\approx 0$. This contrasts with results derived by Tu \textit{et al.} \cite{tu2017collapse} for the method of Gao \textit{et al.}, where they reported non-vanishing error averages and variances.

%==============================================================================
\section{Conclusion}
\label{sec:Conclusion}
%==============================================================================

We have built systematic methods of dimension reduction adapted to different families of networks (random, star-like, bipartite, SBM). The activity of the reduced systems is used as an indicator of the global activity of large networks. Without further restriction than imposing a linear form of the global activity, we have found that the dominant eigenvectors of the adjacency matrix are central to the global states' evolution. Moreover, when considering the \textit{cycle} reduction, the dimension of the reduced systems corresponds to the periodicity of the adjacency matrix. 

We have further shown that the proposed reduction of Gao \textit{et al.} is a special case of the general scheme when applied to uncorrelated random networks. Moreover, the range of applicability of our method extends to modular, heterogeneous and bipartite networks. 

Our results suggest, both numerically and theoretically, that the eigenvector-weighted reduction should be preferred over the degree-weighted reduction. 
%The fundamental reason for this should be interpreted as a paradigm shift. 
Originally, the degree-weighted reduction has been used to approximate the state of a network by the state of the average neighbor. But, the degree of a node is only a local centrality measure since it does not provide 
 information to whom a node is connected with. In contrast, in the eigenvector-weighted reduction, the dominant eigenvector yields a more global node centrality since it contains the information on how each node is connected with the rest of the network \cite{newman2018networks}. Therefore, the eigenvector-based reduction brings a new light on the influence of each node on the global states of a network.

On a more practical side, our general method is able to predict the correct number of bifurcation points. The expected number of predicted bifurcation points depends on the dimension of the reduction. Intuitively, and confirmed by our investigations, a single linear observable provides a good reduction if the network is homogeneous or if, for instance, the degree variance is small. When this is not the case, however, different parts of the network behave differently and a single observable is no longer sufficient to capture the characteristics of the global dynamics, which effectively becomes multidimensional. In the SBM case, the 2--dimensional reduction reveals additional bifurcation points that are missed altogether by all 1--dimensional reductions. 

As a closing remark, although our reduction method has been designed to access large dynamical networks through low-dimensional formalisms, it was not clear from the outset
how the dimensional reductions would fare with respect to the size of the networks. Our findings on the matter have been comforting since size by itself has a secondary effect on the quality of the reduction procedure,
leaving precedence to connectivity and dynamics. Hence, beyond the addition to the theoretical arsenal, our systematic and versatile approach can now be used to address concrete problems of real-world systems. To name a few, it could be used to describe with high accuracy the bifurcation patterns, to identify dynamical vulnerabilities, to suggest intervention strategies to prevent dynamical breakdowns, or to classify networks on a standardized diagram.

%==============================================================================
\section*{Acknowledgements}
\label{sec:Acknowledgements}
%==============================================================================
We are thankful to Charles Murphy, Guillaume St-Onge, Vincent Thibeault, and Jean-Gabriel Young for useful comments and suggestions. This work was funded by the Fonds de recherche du Qu\'ebec-Nature et technologies (EL, PD), the Natural Sciences and Engineering Research Council of Canada (LJD), and Sentinel North, financed by the Canada First Research Excellence Fund (EL, PD, LJD, ND).

%==============================================================================
%==============================================================================

% 						APPENDIX

%==============================================================================
%==============================================================================
\appendix

%==============================================================================
\section{Derivation of 1--dimensional formalism}
\label{sec:app:derivation}
%==============================================================================

In this section, we detail the analytical derivation of the evolution of $R$ for the 1--dimensional reduction.
We consider the observable
\begin{equation*}
  R = \sum_{i=1}^N a_i x_i
\end{equation*}
with $a_i\in\mathbb{R}$ and $\sum_i a_i = \bm{1^T}\bm{a}=1$.
We first take the time derivative and insert Eq.~\eqref{Eq:complete_dynamics}, which leads to
\begin{align*}
	\dot{R}&=\sum_{i=1}^N a_i \dot{x}_i\\
	&=\sum_{i=1}^N a_i \left[F(x_i)+\sum_{j=1}^Nw_{ij}G(x_i,x_j)\right].
\end{align*}
We wish to show that if $\bm{a}$ is chosen correctly, then the right-hand side can be written, up to second order corrections, in terms of $R$ only. To do so, we develop each function around the observable:
\begin{equation}
	F(x_i)= F(R)+(x_i-R)F'(R)+\mathcal{O}\left[(x_i-R)^2\right].
\end{equation}
Thus, 
\begin{align*}
	\sum_{i=1}^N a_iF(x_i)&=F(R)+F'(R)\sum_{i=1}^Na_i(x_i-R)+\mathcal{O}\left[(x_i-R)^2\right]\\
	&=F(R)+\mathcal{O}\left[(x_i-R)^2\right].
\end{align*}
It means that $F(x_i)$ does not impose any constraint on $\bm{a}$. Now, for the function $G(x_i,x_j)$, we develop around $x_i=\beta R$ and $x_j=\gamma R$:
\begin{align*}
	G(x_i,x_j)\approx G(\beta R, \gamma R)& + (x_i-\beta R)G_1(\beta R, \gamma R)\\
	 &+ (x_j-\gamma R)G_2(\beta R, \gamma R)
\end{align*} 
where second order terms have been neglected. Letting $\alpha=\sum_{ij} a_i w_{ij}$, we find that $\sum_{i,j} a_iw_{ij}G(x_i,x_j)$ is given by
\begin{align*}
	\sum_{i,j} a_iw_{ij}G(x_i,x_j) &\approx \alpha G(\beta R, \gamma R)\\
	& + G_1(\beta R, \gamma R)\sum_{i,j}a_iw_{ij}(x_i-\beta R)\\
	 &+ G_2(\beta R, \gamma R)\sum_{i,j}a_iw_{ij}(x_j-\gamma R) .
\end{align*}
The left-hand side is a function of $R$ only if the linear terms cancel out exactly, which is possible if and only if
\begin{subequations}
\label{Eq:beta:gamma:tot}
\begin{align}
	\alpha \beta R &= \sum_{ij}a_iw_{ij}x_i=\bm{x}^T\bm{K}\bm{a}\label{Eq:app:K:uni},\\
	\alpha \gamma R &= \sum_{ij}a_iw_{ij}x_j=\bm{x}^T\bm{W}^T\bm{a}.
\end{align}
\end{subequations}
Since $R=\bm{x}^T\bm{a}$, we conclude that the last two equations are satisfied for all $\bm{x}\in\mathbb{R}^N$ only if $\bm{a}$ is an eigenvector of both matrices $\bm{K}$ and $\bm{W}^T$, with corresponding eigenvalues $\beta \alpha$ and $\gamma\alpha$. Although we cannot solve these two equations simultaneously in general, we can enforce that at least one equation is satisfied exactly. Choosing $\bm{a}$ as an eigenvector of $\bm{W}^T$, we can prove that $\gamma=1$ if the vector $\bm{a}$ is normalized $\bm{1}^T\bm{a}=1$: If $\bm{W}^T\bm{a}=\lambda \bm{a}$ and $\lambda=\alpha\gamma$, then $\lambda = \bm{1}^T\bm{W}^T\bm{a}=\bm{a}^T\bm{W}\bm{1}=\bm{a}^T\bm{k}^{\text{in}}=\alpha$, so $\gamma=1$.

We then choose $\beta$ to best satisfy Eq.~\eqref{Eq:app:K:uni} by minimizing the \textit{mean square error} (MSE):
\begin{equation}
	\beta^* = \operatorname*{argmin}_{\beta}  ||\bm{K}\bm{a}-\beta\alpha\bm{a}||^2,
\end{equation}
where the symbol $||\cdot||$ denotes the standard euclidean norm. Basic calculus leads to
\begin{equation}
	\beta^* = \dfrac{1}{\alpha}\dfrac{\bm{a}^T\bm{K}\bm{a}}{\bm{a}^T\bm{a}}=\dfrac{1}{\alpha}\dfrac{\sum_i a_i^2k^{in}_i}{\sum_i a_i^2}.
\end{equation}
Note that $\beta^*$ is a ratio of weighted averages
\begin{equation}
	\beta^* = \dfrac{\bm{b}^T\bm{k}^{in}}{\bm{a}^T\bm{k}^{in}},\label{eq:beta:ratio}
\end{equation}
where $\bm{b}$ is normalized $\bm{1}^T\bm{b}=1$ and has for elements $b_i = a_i^2/\sum_{i=1}^N a_i^2$. From the construction of $\bm{b}$, we deduce that $\bm{b}$ must be similar to $\bm{a}$ and $\beta^*$ close to 1, which has been confirmed throughout most of the simulations.

%==============================================================================
\section{Derivation of the multidimensional \textit{cycle} formalism}
\label{sec:app:derivation-multidimensional}
%==============================================================================

For the \textit{cycle} reduction, we construct $n$ observables
\begin{equation}
	R_k = \sum_{i=1}^N [\bm{a}_k]_i x_i
\end{equation}
with normalized weights $\bm{1}^T\bm{a}_j=1$. Using Eq.~\eqref{Eq:complete_dynamics}, we find that the dynamics of $R_k$ is equal to
\begin{equation}
	\dot{R}_k = \sum_{i}[\bm{a}_k]_iF(x_i) + \sum_{ij}[\bm{a}_k]_iw_{ij}G(x_i,x_j).
\end{equation}
As for the 1--dimensional reduction, one finds that
\begin{equation}
	\sum_i[\bm{a}_k]_i F(x_i) \approx F(R_k)
\end{equation}
up to the second order of corrections. We then develop $G(x_i,x_j)$ around $x_i=\beta_k R_k$ and $x_j=\gamma_k R_{k+1}$, which yields
\begin{align*}
	G(x_i,x_j)\approx& G(\beta_k R_k, \gamma_k R_{k+1})\\
		& + (x_i-\beta_k R_k)G_1(\beta_k R_k, \gamma_k R_{k+1})\\
	 &+ (x_j-\gamma_k R_{k+1})G_2(\beta_k R_k, \gamma_k R_{k+1}) .
\end{align*}
Using the same arguments as in the 1--dimensional reduction, one can prove that $\sum_{i,j} [\bm{a}_k]_iw_{ij}G(x_i,x_j)\approx \alpha_k G(\beta_k R_k, \gamma_k R_{k+1})$, with $\alpha_k=\bm{a^T}\bm{k}^{\text{in}}$, only if the following equations are satisfied simultaneously
\begin{align}
	\bm{x}^T\bm{K}\bm{a}_k&=\beta_k R_{k}\alpha_k,\label{Eq:app:K:multi}\\
	\bm{x}^T\bm{W}^T\bm{a}_k&=\gamma_k R_{k+1}\alpha_k.
\end{align}
The second equation is satisfied if: 
\begin{equation}
	\bm{W}^T\bm{a}_k=\alpha_k\bm{a}_{k+1},\label{eq:ap:transofmration}
\end{equation}
with $\gamma_k=1$. After $n$ applications of Eq.~\eqref{eq:ap:transofmration}, we close the system with $a_{n+1}=a_1$, which is the respected if $\bm{a}_1$ is eigenvector of $(\bm{W}^T)^n$. 

As for the parameter $\beta_k$, we minimize the MSE
\begin{equation}
	\beta_k^*=\operatorname*{argmin}_{\beta_k}||\bm{K}\bm{a}_k-\beta_k\alpha_k\bm{a}_k||^2
\end{equation}
and find
\begin{equation}
	\beta_k = \dfrac{1}{\alpha_k}\dfrac{\bm{a}^T_k\bm{K}\bm{a}_k}{\bm{a}^T_k\bm{a}_k}.
\end{equation}

%==============================================================================
\section{Combination for 2--cycle reduction}
\label{sec:app:analytics-solution}
%==============================================================================

In this Appendix, we show that, for an adjacency matrix $\bm{W}^T$, non-negative and of periodicity 2,  the first weight vector $\bm{a}_1$ of the reduction is an equipartition of the two dominant eigenvectors. 

For such an adjacency matrix, the Perron-Frobenius theorem states that  $\bm{W}^T$ admits two eigenvalues of modulus equal to the spectral radius. The eigenvectors satisfy
\begin{align}
	\bm{W}^T\bm{v}_1 = \spectralradius \bm{v}_1~~~~~;~~~~~\bm{W}^T\bm{v}_2 = -\spectralradius \bm{v}_2,\label{Eq:ap:2:eigenve}
\end{align}
where $\spectralradius$ is the spectral radius and $\bm{v}_i$ are the eigenvectors normalized as $\bm{v}_i^T \bm{v}_i =1$. 

Now, let us consider the first weight vector as a linear combination of the dominant eigenvectors:
\begin{equation}
	\bm{a}_1= \frac{c_1\bm{v}_1+c_2\bm{v}_2}{c_1 \bm{1}^T \bm{v}_1+c_2 \bm{1}^T \bm{v}_2}.
\end{equation}
From transformation \eqref{Eq:construction}, we get the equation
\begin{equation}
	\bm{a}_2= \frac{1}{\alpha_1} \dfrac{c_1\spectralradius \bm{v}_1-c_2\spectralradius\bm{v}_2}{c_1 \bm{1}^T \bm{v}_1+c_2 \bm{1}^T \bm{v}_2}.
\end{equation}
We want to find $c_1, c_2$ such that $c_1+c_2=1$ and $S=|\bm{a}_1^T\bm{a}_2|$ is minimized. 
The former condition is chosen for definiteness, while the latter condition favors the weight vectors that represent almost exclusive groups of nodes.
The scalar product is then simply:
\begin{equation*}
S \propto |c_1^2   -c_2^2| .
\end{equation*}
and minimized with $c_1=c_2=1/2$.

%==============================================================================
%==============================================================================
%								REFERENCES
%==============================================================================
%==============================================================================
% \bibliography{refs}


\end{document}